\documentclass{sail-report-class}

\reportnumber{3}

\title{When Stress Becomes Signal: Detecting Antifragility-Compatible Regimes in Multi-Agent LLM Systems}
\author{%
  Jose Manuel de la Chica\textsuperscript{1, $\dagger$}\,\orcidlink{0000-0003-4567-8901} \quad
  Juan Manuel Vera\textsuperscript{1}\,\orcidlink{0000-0002-3456-7890} \quad
  Jairo Rodíguez\textsuperscript{1}\,\orcidlink{0000-0003-4567-8901} \quad
  \normalsize\textsuperscript{1}Santander AI Lab
}
\correspondingauthor{josema.chica@gruposantander.com}
\date{\today}

\begin{document}

\begin{abstract}
Multi-agent LLM systems are increasingly used to solve complex tasks through
decomposition, debate, specialization, and ensemble reasoning. However, these
systems are usually evaluated in terms of robustness: whether performance is
preserved under perturbation. This paper studies a different question: whether
semantic stress exposes structured variation that could support future
antifragile learning. We introduce CAFE (Cognitive Antifragility Framework for Evaluation), a statistical framework for detecting
antifragility-compatible regimes in multi-agent architectures. CAFE models a
controlled expected distribution of semantic stressors, reconstructs an
architecture-specific observed effective stress distribution from
multi-dimensional judge signals, and compares both distributions using a
distributional Jensen Gap under a convex stress potential. A positive gap does
not imply immediate performance improvement; instead, it indicates a
convex-expansive deformation of the observed stress distribution, suggesting
that the architecture exposes learnable stress structure. We evaluate CAFE on a
banking-risk analysis benchmark with five multi-agent architectures: flat,
hierarchical, debate, meta-adaptive, and ensemble. Across all architectures,
semantic stress reduces average judged quality by roughly one third. Yet all
architectures exhibit positive distributional Jensen Gaps with bootstrap
confidence intervals above zero. These results show that immediate quality
degradation can coexist with statistically detectable
antifragility-compatible stress geometry. CAFE is therefore not an
antifragile learner itself, but a measurement layer for identifying when and
where antifragility learning may be worth applying.
\end{abstract}

\maketitle

\section{Introduction}
\label{sec:introduction}

Large language models are increasingly deployed as multi-agent systems: flat
pipelines, specialist hierarchies, adversarial debates, ensembles, and
meta-controlled workflows. These architectures are attractive because complex
tasks often require decomposition, disagreement, synthesis, and adaptation.
However, real inputs are rarely clean. In high-stakes analytical domains, such
as financial risk assessment, prompts may contain contradictory evidence,
overloaded context, ambiguous references, and temporally stale information.
These conditions are not peripheral noise; they are part of the operating
environment.

Most evaluation protocols ask whether a model or agentic architecture remains
robust under such perturbations. Robustness is important, but it is not the
same as antifragility. A robust system preserves performance under stress. A
fragile system degrades. An antifragile system, in the stronger sense, should
eventually benefit from adversity. For current LLM-based multi-agent systems,
claiming immediate improvement under harder inputs is usually too strong:
semantic stress often lowers answer quality. The more useful question is
whether stress exposes structured variation that a future adaptive mechanism
could exploit.

This paper studies that intermediate regime. We ask whether a multi-agent
architecture, when exposed to a known distribution of semantic stressors,
produces an observed stress-response geometry that is statistically compatible
with future antifragile learning. We call such a regime
\emph{antifragility-compatible}: the system is not necessarily improving yet,
but its response to stress contains structured, convex-expansive variation
rather than mere collapse or noise.

We propose CAFE, a statistical framework for detecting this regime. CAFE starts
from a controlled expected stress distribution over four semantic stress
dimensions: conflict, load, ambiguity, and temporal drift. Each architecture is
then evaluated with a multi-dimensional judge that measures coherence,
grounded novel inference, contradiction resolution, and structural
preservation. A multi-output polynomial response model maps designed stress
intensities to judge response signals. We then solve an inverse reconstruction
problem to estimate the architecture-specific observed effective stress
distribution. Finally, CAFE compares the expected and observed stress
distributions through a distributional Jensen Gap under a convex stress
potential.

The resulting statistic is not a Jensen Gap over task quality. Instead, it
measures whether the observed effective stress distribution expands or
compresses relative to the expected stress distribution. A positive gap
indicates a convex-expansive deformation, which we interpret as an
antifragility-compatible opportunity: the architecture exposes learnable stress
structure. A near-zero gap indicates resilience, and a negative gap indicates
fragile compression.

We evaluate CAFE on a controlled banking-risk analysis benchmark with five
multi-agent architectures: a flat baseline, a hierarchical specialist system,
an adversarial debate system, a meta-adaptive controller, and an ensemble. The
main empirical finding is intentionally two-sided. All architectures lose
average judged quality under stress, with relative quality drops of roughly one
third. Nevertheless, all architectures exhibit positive distributional Jensen
Gaps with bootstrap confidence intervals above zero. This shows that immediate
quality degradation can coexist with statistically detectable
antifragility-compatible stress geometry.

Our contributions are:
\begin{itemize}
    \item We formalize antifragility-compatible evaluation for multi-agent LLM
    architectures as a distributional deformation problem rather than as
    immediate performance improvement.
    \item We introduce a multi-output polynomial response model for
    reconstructing observed effective stress from judge response signals.
    \item We define a distributional Jensen Gap that compares expected and
    observed stress distributions under a convex stress potential.
    \item We evaluate five multi-agent architectures under controlled semantic
    stress and show that quality degradation can coexist with positive
    antifragility-compatible stress geometry.
\end{itemize}

CAFE is therefore not itself an antifragile learner. It is a measurement layer:
it identifies when and where antifragility learning may be worth applying. This
distinction matters because adaptive systems need reliable signals before they
can learn from adversity rather than merely endure it.

\section{Related Work}
\label{sec:related-work}

\subsection*{Antifragility and convex response.}
Antifragility distinguishes systems that merely tolerate stress from systems
that can extract useful structure from volatility, uncertainty, or perturbation.
Taleb and Douady formalize antifragility through nonlinear exposure to
dispersion and model error, connecting beneficial stress response to convexity
\cite{taleb2013mathematical}. Related work further develops the link between
convex response, fragility, and beneficial variation
\cite{taleb2013antifragility,taleb2023convex}. Systems and software research
has adapted this idea to systems-of-systems, software architectures, and cloud
settings \cite{johnson2013antifragility,russo2017towards,botros2024towards}.
These works typically study operational stress such as failures, latency,
recovery, or chaos-engineering perturbations. CAFE instead studies semantic and
epistemic stress in language-agent architectures. Its goal is not to claim that
an architecture immediately improves under adversity, but to detect whether the
observed stress-response geometry is statistically compatible with future
antifragile adaptation.

\subsection*{Response surfaces and distributional comparison.}
CAFE uses response-surface modeling to estimate how judge signals vary as a
function of controlled stress intensities. This connects to classical
computer-experiment methodology, where expensive systems are probed at
controlled input settings and approximated by interpretable surrogate models
\cite{sacks1989design}. Our use of a polynomial model is not the final
antifragility test; it is an instrument for reconstructing an
architecture-specific observed stress distribution. The final statistic compares
the designed and observed stress distributions through a distributional Jensen
Gap. This places CAFE near distributional comparison methods such as
Jensen-Shannon divergence \cite{lin1991divergence}, maximum mean discrepancy
\cite{gretton2012kernel}, and energy-distance statistics
\cite{szekely2013energy}. Unlike these generic distances, CAFE asks whether
the observed distribution expands under a convex stress potential, revealing a
regime that could be exploited by antifragile learning.

\subsection*{LLM agents and coordination.}
Recent LLM systems increasingly use multiple agents with specialized roles,
communication protocols, and coordination mechanisms. AutoGen provides a
general framework for multi-agent LLM conversations \cite{wu2023autogen};
CAMEL studies role-playing agents as a substrate for cooperative behavior
\cite{li2023camel}; and MetaGPT encodes structured workflows into
multi-agent collaboration \cite{hong2023metagpt}. Surveys of LLM-based
multi-agent systems organize this space around profiling, communication,
planning, cooperation, and evaluation \cite{guo2024large}. Our experimental
architectures instantiate common coordination patterns: a flat baseline,
hierarchical decomposition, debate, meta-adaptive control, and independent
ensemble synthesis. Prior work shows that multi-agent debate can improve
reasoning and factuality \cite{du2024improving} and that sampling multiple
reasoning paths can improve answers through self-consistency
\cite{wang2023selfconsistency}. CAFE complements these results by asking a
different question: which coordination patterns expose structured stress
variation that an adaptive system could later learn from?

\subsection*{Reflection, adaptation, and learning from adversity.}
Several agentic methods improve outputs through feedback, reflection, search,
or memory. Self-Refine iteratively revises outputs using self-feedback
\cite{madaan2023selfrefine}; Reflexion stores verbal feedback to improve
future trials \cite{shinn2023reflexion}; Tree of Thoughts searches over
intermediate reasoning states \cite{yao2023tree}; and ReAct interleaves
reasoning with action \cite{yao2023react}. These methods demonstrate that LLM
systems can use additional signals to adapt at inference time or across trials.
However, they usually measure adaptation by final task success or preference.
CAFE targets the condition that precedes such adaptation: whether stress
produces structured, convex-expansive variation in the observed stress space.
In this sense, CAFE is a detector of when antifragile learning may be worth
applying.

\subsection*{Stress testing and holistic evaluation.}
Average benchmark accuracy is insufficient for evaluating language systems
under distribution shift. CheckList introduced behavioral tests for targeted
NLP capabilities \cite{ribeiro2020checklist}; Robustness Gym unified
subpopulation, transformation, and adversarial evaluations
\cite{goel2021robustnessgym}; Dynabench proposed dynamic model-in-the-loop
benchmarking \cite{kiela2021dynabench}; and HELM argues for multi-metric
evaluation across scenarios and desiderata \cite{liang2022helm}. Our stress
dimensions connect to benchmarks for truthfulness and verification
\cite{thorne2018fever,lin2022truthfulqa}, ambiguity
\cite{min2020ambigqa}, long-context stress \cite{bai2023longbench}, and
automated red-teaming \cite{perez2022red}. CAFE differs by treating stress as
a continuous random variable and by measuring the distributional deformation
from expected to observed stress, rather than only reporting accuracy on a
fixed challenge set.

\subsection*{LLM-as-judge evaluation.}
Open-ended multi-agent responses require structured evaluation beyond
exact-match metrics. LLM-as-judge methods such as G-Eval use explicit rubrics
to align automatic evaluation with human judgment \cite{liu2023geval}, while
MT-Bench and Chatbot Arena study the scalability and biases of judge-based
evaluation \cite{zheng2023judging,chiang2024chatbot}. CAFE uses an explicit
judge as a measurement component that returns multiple response signals:
coherence, grounded inference, contradiction resolution, and structural
preservation. These scores are not treated as a final scalar performance
measure. Instead, they constrain the reconstruction of an observed effective
stress distribution, which is then compared to the expected stress distribution
through the CAFE Jensen Gap.

\section{Method}
\label{sec:method}

CAFE is a statistical framework for detecting whether a multi-agent
architecture is operating in an antifragility-compatible regime. The framework
does not define antifragility as immediate improvement in average output
quality. Instead, it asks whether stress induces a structured,
convex-expansive deformation between the expected stress distribution and the
observed effective stress distribution. Such a regime is valuable because it
exposes variation that future antifragile learning, routing, memory, or
architecture-selection mechanisms could exploit.

\subsection{Problem Formulation}
\label{subsec:problem-formulation}

Let \(\mathcal{A}\) be a set of candidate agentic architectures. For an
architecture \(a\in\mathcal{A}\), a clean prompt \(u\), and a stress vector
\begin{equation}
    \boldsymbol{\psi}
    =
    (\psi_1,\psi_2,\psi_3,\psi_4)
    \in [0,1]^4,
\end{equation}
a perturbation operator \(\tau\) generates a stressed prompt
\begin{equation}
    \tilde{u} = \tau(u,\boldsymbol{\psi}).
\end{equation}
The architecture produces an output
\begin{equation}
    y_a = a(\tilde{u}).
\end{equation}
An evaluator \(J\) maps the prompt-output pair into a vector of response
signals
\begin{equation}
    \mathbf{s}_{a}
    =
    J(\tilde{u},y_a)
    \in [0,1]^K.
\end{equation}
In our experiments \(K=4\), corresponding to coherence, grounded novel
inference, contradiction resolution, and structural preservation. We keep these
signals as a vector rather than collapsing them into a single quality score,
because the inverse reconstruction of observed stress is better constrained by
multiple response measurements. The empirical dataset for architecture \(a\) is
\begin{equation}
    \mathcal{D}_a
    =
    \{(\boldsymbol{\psi}_n,\mathbf{s}_{a,n})\}_{n=1}^{N_a}.
\end{equation}

\subsection{Stress Space}
\label{subsec:stress-space}

The stress vector contains four semantic stress dimensions:
\begin{equation}
    \boldsymbol{\psi}
    =
    (
    \psi_{\mathrm{conflict}},
    \psi_{\mathrm{load}},
    \psi_{\mathrm{ambiguity}},
    \psi_{\mathrm{drift}}
    ).
\end{equation}
These dimensions encode contradictory evidence, semantic overload, structural
or referential ambiguity, and temporal inconsistency. We center stress
coordinates as
\begin{equation}
    x_i = \psi_i - \frac{1}{2},
    \qquad
    x_i\in[-1/2,1/2],
\end{equation}
and write \(\mathbf{x}\) for the centered stress vector.

\subsection{Multi-Output Response Model}
\label{subsec:multi-output-response-model}

For each architecture \(a\), we assume a latent response-signal surface
\begin{equation}
    \mathcal{S}_a : [-1/2,1/2]^4 \rightarrow [0,1]^K,
\end{equation}
with observations
\begin{equation}
    \mathbf{s}_{a,n}
    =
    \mathcal{S}_a(\mathbf{x}_n)
    +
    \boldsymbol{\varepsilon}_{a,n},
    \qquad
    \mathbb{E}[\boldsymbol{\varepsilon}_{a,n}\mid\mathbf{x}_n]=\mathbf{0}.
\end{equation}
Each response dimension is approximated with the same interpretable polynomial
basis. For judge dimension \(k\), we fit
\begin{align}
    \widehat{\mathcal{S}}_{a,k}(\mathbf{x})
    =&\;
    \theta_{0,a,k}
    + \sum_{i=1}^{4}\theta_{i,a,k}x_i
    + \sum_{i=1}^{4}\alpha_{i,a,k}x_i^2
    + \sum_{1\leq i<j\leq4}\gamma_{ij,a,k}x_ix_j
    \nonumber\\
    &\;
    + \delta_{12,a,k}x_1x_2^2
    + \delta_{34,a,k}x_3x_4^2.
    \label{eq:polynomial-response}
\end{align}
The linear terms estimate direct stress sensitivity, the quadratic terms
estimate marginal curvature, the bilinear terms capture cross-stressor
interference, and the two selected cubic terms allow conflict to modulate the
effect of load and ambiguity to modulate the effect of drift. The full
multi-output predictor is
\begin{equation}
    \widehat{\mathcal{S}}_a(\mathbf{x})
    =
    (
    \widehat{\mathcal{S}}_{a,1}(\mathbf{x}),\ldots,
    \widehat{\mathcal{S}}_{a,K}(\mathbf{x})
    ).
\end{equation}

Coefficients are estimated with ridge-regularized least squares:
\begin{equation}
    \widehat{\boldsymbol{\beta}}_{a,k}
    =
    \arg\min_{\boldsymbol{\beta}}
    \sum_{n=1}^{N_a}
    \left(
    s_{a,n,k}
    -
    \mathcal{S}_{\boldsymbol{\beta}}(\mathbf{x}_n)
    \right)^2
    +
    \rho\lVert\boldsymbol{\beta}_{-0}\rVert_2^2,
    \label{eq:coefficient-estimation}
\end{equation}
where the intercept is excluded from the penalty. This model is not itself the
antifragility criterion. It is an instrument for reconstructing the effective
stress profile that best explains the observed response signals.

\subsection{Observed Effective Stress Reconstruction}
\label{subsec:observed-effective-stress}

All architectures are exposed to the same designed stress distribution
\(P_{\Psi}\), or \(P_X\) in centered coordinates. However, different
architectures can transform the same designed stress into different observed
effective stress distributions. Given the observed judge vector
\(\mathbf{s}_{a,n}\), we reconstruct the effective observed stress by solving
\begin{equation}
    \widehat{\mathbf{x}}^{\mathrm{obs}}_{a,n}
    =
    \arg\min_{\mathbf{x}\in[-1/2,1/2]^4}
    \left\|
    \mathbf{s}_{a,n}
    -
    \widehat{\mathcal{S}}_a(\mathbf{x})
    \right\|_W^2
    +
    \lambda R(\mathbf{x}),
    \label{eq:inverse-stress-estimation}
\end{equation}
where \(\lVert\mathbf{z}\rVert_W^2=\mathbf{z}^\top W\mathbf{z}\). The
regularizer resolves local non-identifiability and stabilizes the inverse. We
use either a distributional prior,
\begin{equation}
    R(\mathbf{x})
    =
    (\mathbf{x}-\boldsymbol{\mu}_X)^\top
    \Sigma_X^{-1}
    (\mathbf{x}-\boldsymbol{\mu}_X),
\end{equation}
or an anchored prior,
\begin{equation}
    R(\mathbf{x})=\lVert\mathbf{x}-\mathbf{x}_n\rVert_2^2.
\end{equation}
The reconstructed stress vector in the original scale is
\begin{equation}
    \widehat{\boldsymbol{\psi}}^{\mathrm{obs}}_{a,n}
    =
    \widehat{\mathbf{x}}^{\mathrm{obs}}_{a,n}
    +
    \frac{1}{2}\mathbf{1}.
\end{equation}
The resulting empirical observed stress distribution is
\begin{equation}
    \widehat{P}^{\mathrm{obs}}_{\Psi,a}
    =
    \frac{1}{N_a}
    \sum_{n=1}^{N_a}
    \delta_{\widehat{\boldsymbol{\psi}}^{\mathrm{obs}}_{a,n}}.
\end{equation}

\subsection{Distributional Deformation}
\label{subsec:distributional-deformation}

CAFE studies the deformation from expected to observed stress. For each
architecture, define a map
\begin{equation}
    T_a:[-1/2,1/2]^4\rightarrow[-1/2,1/2]^4,
    \qquad
    \widehat{\mathbf{x}}^{\mathrm{obs}}_{a,n}
    \approx
    T_a(\mathbf{x}_n).
\end{equation}
The push-forward \(T_a\#P_X\) is the architecture-specific observed stress law.
To evaluate whether this deformation is expansive under a convex notion of
stress magnitude, we introduce a convex potential
\begin{equation}
    \phi:[-1/2,1/2]^4\rightarrow\mathbb{R}_+.
\end{equation}
In the experiments we use \(\phi(\mathbf{x})=\lVert\mathbf{x}\rVert_2^2\), so
the statistic measures total dispersion around the mean stress vector. Other
convex potentials can emphasize specific stress dimensions.

\subsection{Distributional Jensen Gap}
\label{subsec:jensen-gap}

The expected Jensen dispersion is
\begin{equation}
    \mathcal{V}^{\mathrm{exp}}
    =
    \mathbb{E}_{X\sim P_X}[\phi(X)]
    -
    \phi(\mathbb{E}_{X\sim P_X}[X]).
\end{equation}
The observed Jensen dispersion for architecture \(a\) is
\begin{equation}
    \mathcal{V}^{\mathrm{obs}}_a
    =
    \mathbb{E}_{X\sim P_X}[\phi(T_a(X))]
    -
    \phi(\mathbb{E}_{X\sim P_X}[T_a(X)]).
\end{equation}
The CAFE distributional Jensen Gap is
\begin{equation}
    \mathcal{G}_a
    =
    \mathcal{V}^{\mathrm{obs}}_a
    -
    \mathcal{V}^{\mathrm{exp}}.
    \label{eq:distributional-jensen-gap}
\end{equation}
A positive \(\mathcal{G}_a\) indicates a convex-expansive deformation of the
observed stress distribution. We call this an antifragility-compatible regime:
the system is not necessarily improving yet, but the stress response exposes
structured variation that can be used by later adaptation. A gap near zero
indicates resilience, and a negative gap indicates compression of observed
stress variability.

\section{Experimental Setup}
\label{sec:experimental-setup}

We evaluate CAFE on a controlled banking-risk analysis task. The experiment is
designed to answer a specific question: when multi-agent architectures are
exposed to semantic stress, do their responses reveal an
antifragility-compatible stress geometry that could later be exploited by
adaptive learning?

\subsection{Synthetic Stress Dataset}
\label{subsec:synthetic-stress-dataset}

The task domain is banking-risk assessment. Each clean prompt asks an
architecture to assess systemic risk for a financial institution using
indicators such as CET1 ratio, liquidity coverage ratio, non-performing loans,
trading VaR, operational losses, wholesale funding gap, interest-rate
conditions, and mitigation recommendations. The domain is useful because it
naturally combines multiple interacting risk channels and requires explicit
handling of conflicting, stale, overloaded, or ambiguous information.

We generate \(50\) clean prompts and \(10\) perturbed variants per clean prompt,
for \(500\) stressed prompts. Stress intensities are sampled independently from
Gaussian distributions and clipped to \([0,1]\):
\begin{equation}
    \psi_i\sim \mathrm{clip}_{[0,1]}(\mathcal{N}(\mu_i,\tau_i^2)).
\end{equation}
\begin{center}
\begin{tabular}{lccp{7.2cm}}
\hline
Stress dimension & Mean & Std. & Perturbation operator \\
\hline
\(\psi_{\mathrm{conflict}}\) & 0.40 & 0.18 & Contradictory banking-risk claims. \\
\(\psi_{\mathrm{load}}\) & 0.58 & 0.20 & Extra caveats, ratios, dependencies, and supervisory notes. \\
\(\psi_{\mathrm{ambiguity}}\) & 0.32 & 0.16 & Ambiguous entities, metrics, and references. \\
\(\psi_{\mathrm{drift}}\) & 0.46 & 0.18 & Stale or temporally inconsistent reporting windows. \\
\hline
\end{tabular}
\end{center}
The generation seed is \(20260428\). Clean prompts are retained as the reference
condition, while perturbed prompts define the expected stress distribution.

\subsection{Architectures}
\label{subsec:agentic-architectures}

We evaluate five architectures with different coordination mechanisms. All
architectures receive the same prompts and are instructed to preserve
banking-risk terminology, reason over the stressed input as given, explicitly
handle contradiction, overload, ambiguity, and drift, and produce concrete risk
recommendations.

\begin{center}
\begin{tabular}{llll}
\hline
ID & Architecture & Coordination & Expected calls \\
\hline
A0 & Flat Pipeline & none & 1 \\
A1 & Hierarchical & top-down delegation & 5 \\
A2 & Adversarial Debate & thesis--antithesis--synthesis & 3 \\
A3 & Meta-Adaptive & feedback-driven switching & \(4+\) \\
A4 & Simple Ensemble & independent diversity and consensus & 4 \\
\hline
\end{tabular}
\end{center}

\begin{figure}[t]
    \centering
    \begin{minipage}{0.32\linewidth}
        \centering
        \includegraphics[width=\linewidth]{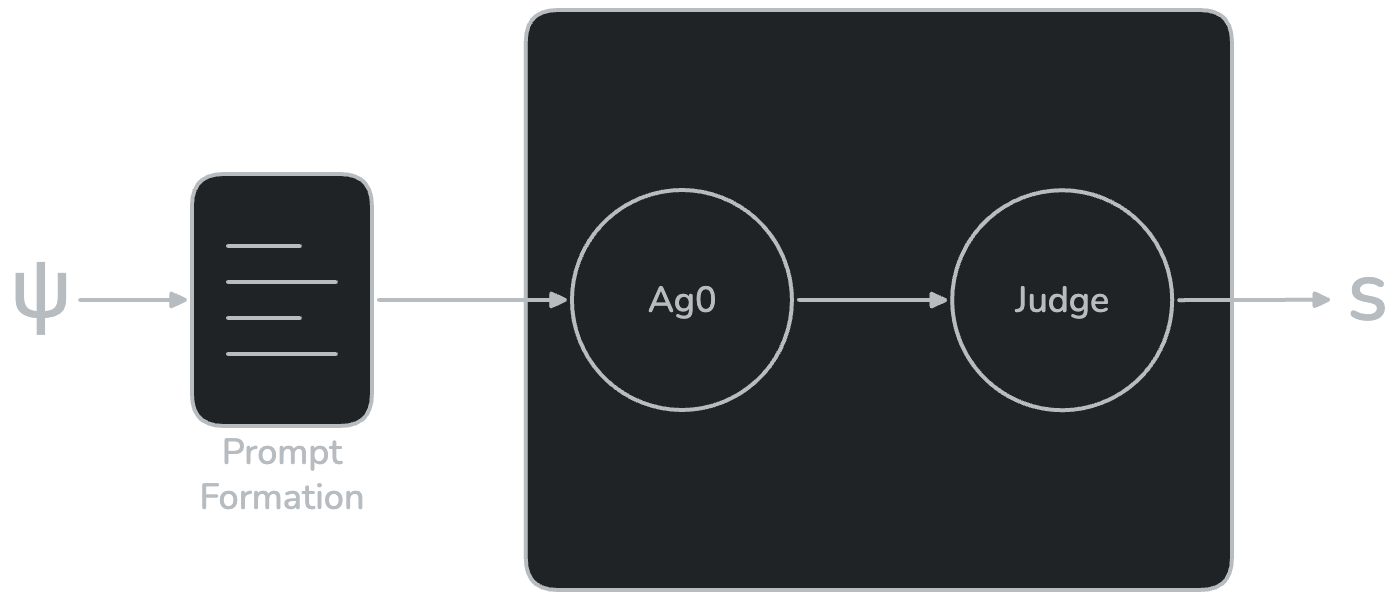}\\
        \small A0: Flat Pipeline
    \end{minipage}
    \hfill
    \begin{minipage}{0.32\linewidth}
        \centering
        \includegraphics[width=\linewidth]{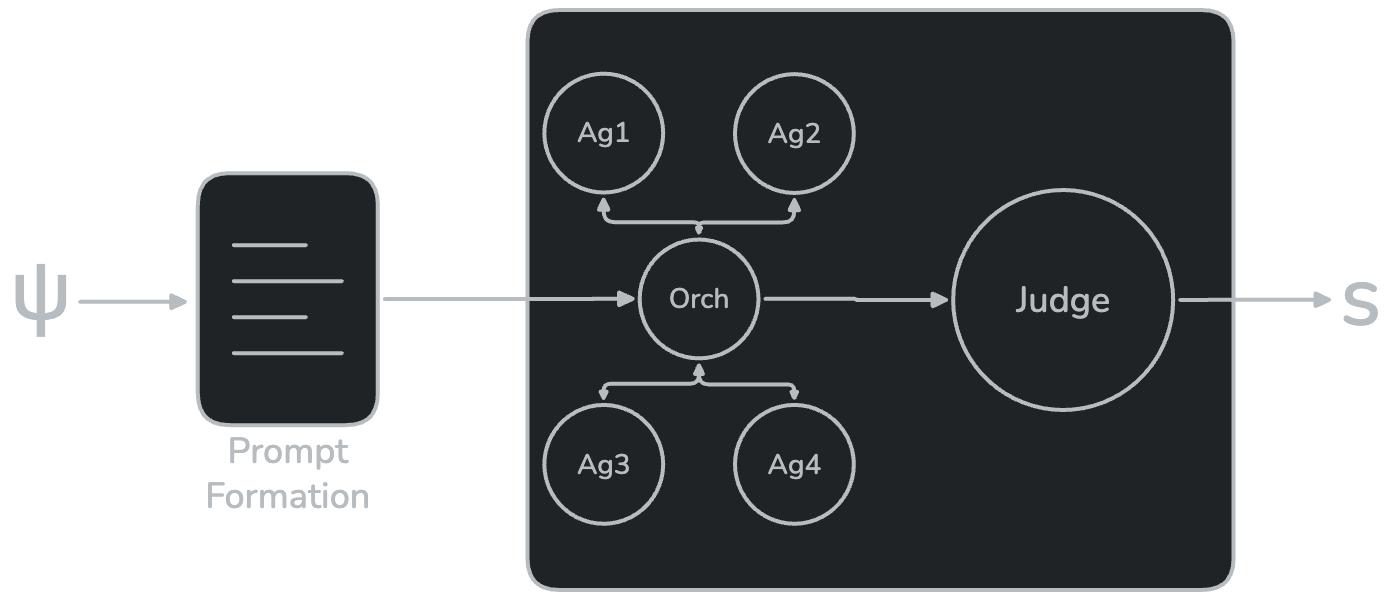}\\
        \small A1: Hierarchical
    \end{minipage}
    \hfill
    \begin{minipage}{0.32\linewidth}
        \centering
        \includegraphics[width=\linewidth]{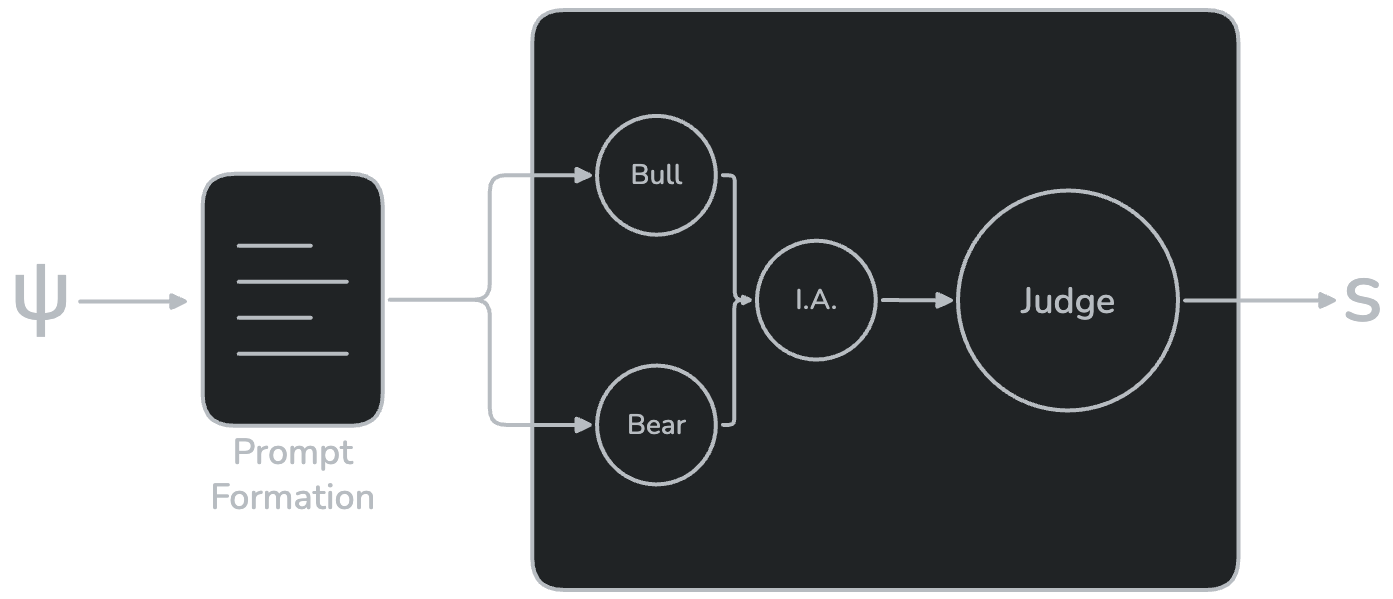}\\
        \small A2: Debate
    \end{minipage}

    \vspace{0.8em}

    \begin{minipage}{0.40\linewidth}
        \centering
        \includegraphics[width=\linewidth]{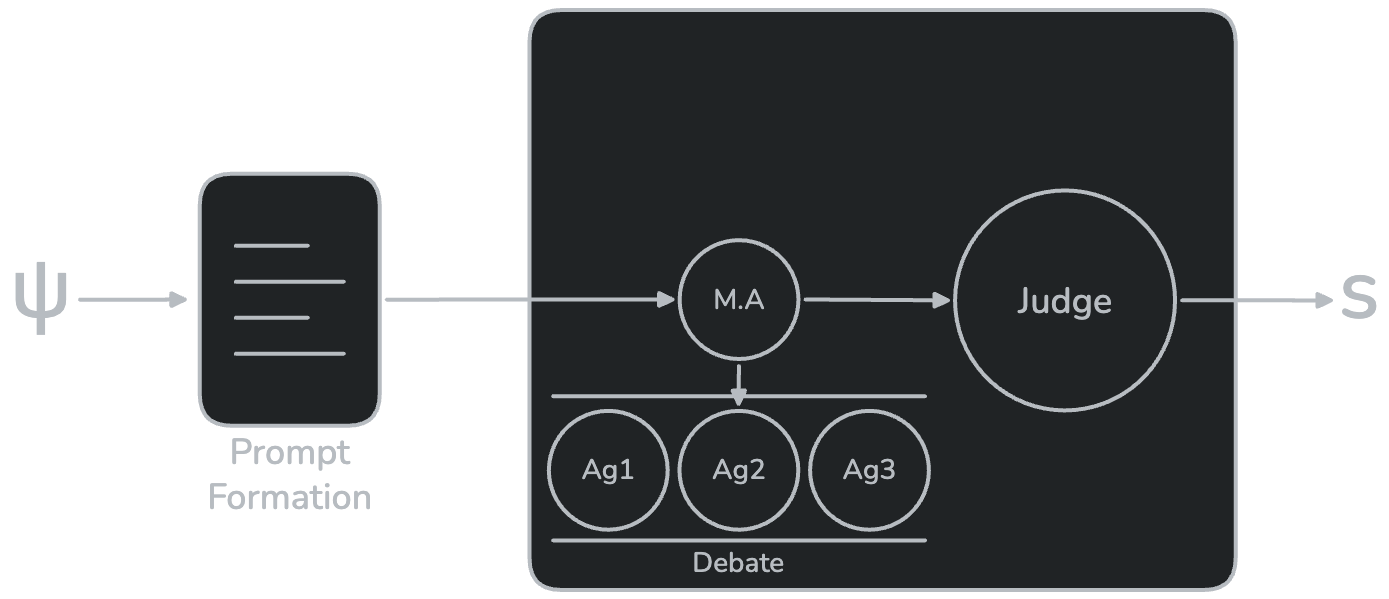}\\
        \small A3: Meta-Adaptive
    \end{minipage}
    \hspace{0.08\linewidth}
    \begin{minipage}{0.40\linewidth}
        \centering
        \includegraphics[width=\linewidth]{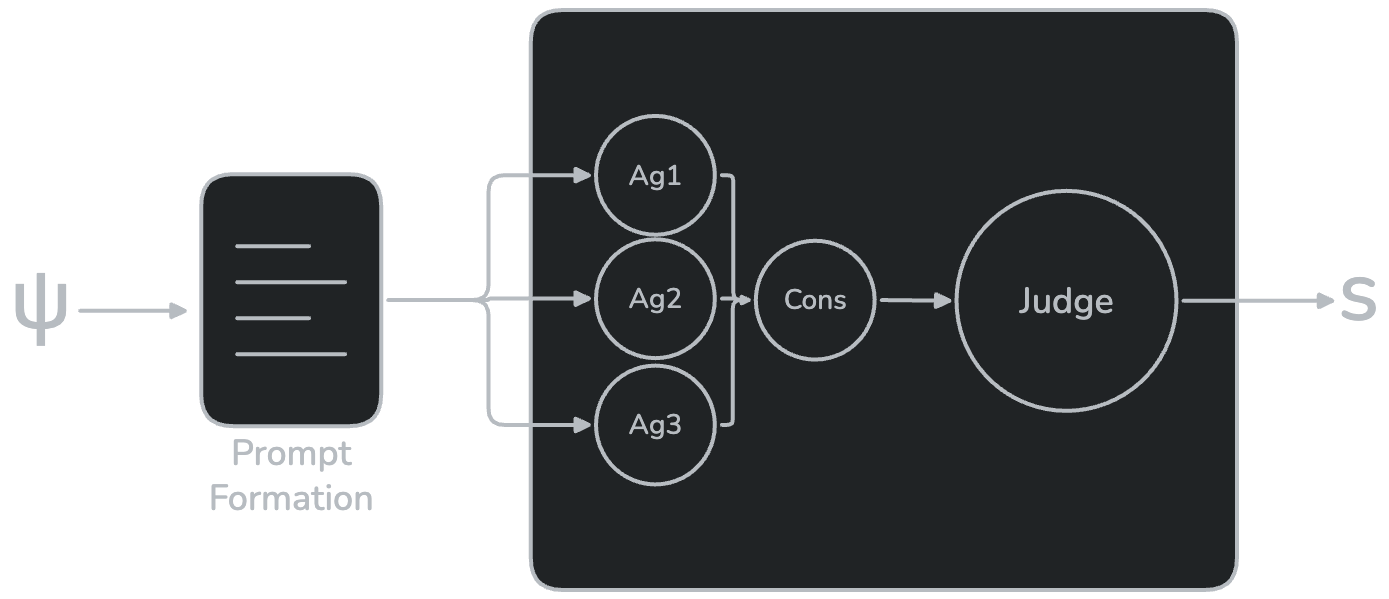}\\
        \small A4: Ensemble
    \end{minipage}
    \caption{Agentic architectures evaluated in CAFE.}
    \label{fig:agentic-architectures}
\end{figure}

Figure \ref{fig:agentic-architectures} shows implemented architectures where A0 is a single-agent baseline. A1 decomposes the problem into credit, market,
liquidity, and operational-risk specialists before synthesis. A2 uses a risk
optimist (bull), a risk pessimist (bear), and an arbitrator. A3 begins with debate and adds a
meta-controller that can request stricter synthesis when coherence or drift
degrades. A4 runs three independent analysts and aggregates them through a
consensus builder.

\subsection{Judge and Response Signals}
\label{subsec:agentic-judge}

Each response is scored by a deterministic judge using four dimensions:
\begin{equation}
    \mathbf{s}_{a,n}
    =
    (
    s^{\mathrm{coh}}_{a,n},
    s^{\mathrm{inf}}_{a,n},
    s^{\mathrm{contr}}_{a,n},
    s^{\mathrm{struct}}_{a,n}
    )
    \in[0,1]^4.
\end{equation}
The dimensions are coherence, novel inference rate, contradiction resolution,
and structural drift inverse. The judge uses only the input prompt and the
system output. It penalizes unsupported claims, ignored contradiction, ignored
temporal drift, ignored overload, and overconfident handling of ambiguity. The
four scores are used as response signals for reconstructing observed effective
stress; they are not treated as a single final performance metric in the CAFE
criterion.

\subsection{Modeling and Reconstruction}
\label{subsec:statistical-modeling-pipeline}

For each architecture, clean and perturbed samples are merged into a modeling
table. A fully evaluated architecture contributes \(550\) rows: \(50\) clean and
\(500\) perturbed samples. We center stress intensities as \(x_i=\psi_i-1/2\)
and fit the multi-output polynomial model from Section~\ref{sec:method} with
ridge regularization \(\rho=10^{-3}\). We report \(R^2\), RMSE, and MAE for
the fitted response-signal surfaces.

Observed effective stress is reconstructed with the inverse problem in
Equation~\eqref{eq:inverse-stress-estimation}. We use an anchored regularizer,
inverse weight \(\lambda=0.05\), L-BFGS-B optimization, bounds
\([-1/2,1/2]^4\), and at most \(100\) iterations. We then fit an
architecture-specific deformation map \(T_a\) from designed to reconstructed
stress using the same polynomial basis.

\subsection{Distributional Jensen Evaluation}
\label{subsec:distributional-jensen-evaluation}

We compute the expected Jensen dispersion from the designed stress vectors and
the observed Jensen dispersion from the reconstructed observed stress vectors.
The convex potential is \(\phi(\mathbf{x})=\lVert\mathbf{x}\rVert_2^2\), so the
gap measures expansion or compression of total stress dispersion around the
mean. We classify gaps using a resilience tolerance of \(0.01\): values above
\(0.01\) are antifragility-compatible, values below \(-0.01\) are fragile, and
intermediate values are resilient. Uncertainty is estimated with \(500\)
bootstrap resamples using seed \(7\), and \(2.5\%\) and \(97.5\%\) quantiles are
reported as confidence intervals.

\section{Results}
\label{sec:results}

We report results for A0 Flat, A1 Hierarchical, A2 Debate, A3 Meta-Adaptive,
and A4 Ensemble. The main empirical finding is deliberately two-sided: average
judge quality decreases under stress for every architecture, yet every
architecture exhibits a positive distributional Jensen Gap. CAFE therefore
detects antifragility-compatible regimes, not automatic improvement under
adversity.

\subsection{Quality Drops Under Perturbation}
\label{subsec:quality-decreases}

Table~\ref{tab:quality-drop} reports mean judge quality on clean and perturbed
prompts, computed as the average of the four judge dimensions. All
architectures degrade under perturbation. The largest relative drop is observed
for A4 Ensemble, despite A4 later showing the strongest CAFE gap.

\begin{table}[t]
\centering
\caption{Mean judge quality on clean and perturbed prompts. CAFE separates this
immediate quality drop from the distributional opportunity for antifragile
adaptation.}
\label{tab:quality-drop}
\begin{tabular}{lrrrr}
\hline
Architecture & Clean quality & Perturbed quality & Drop & Relative drop \\
\hline
A0 Flat & 0.8309 & 0.5663 & -0.2646 & 31.8\% \\
A1 Hierarchical & 0.8139 & 0.5473 & -0.2665 & 32.8\% \\
A2 Debate & 0.8339 & 0.5527 & -0.2812 & 33.7\% \\
A3 Meta-Adaptive & 0.8206 & 0.5532 & -0.2674 & 32.6\% \\
A4 Ensemble & 0.8400 & 0.5426 & -0.2974 & 35.4\% \\
\hline
\end{tabular}
\end{table}

This result rules out a naive interpretation of antifragility as immediate
performance improvement. The architectures do not become better simply because
the prompts are stressed. The relevant question is whether stress reveals
structured variation that an adaptive mechanism could learn from.

\subsection{Positive Jensen Gaps Reveal Antifragility-Compatible Regimes}
\label{subsec:distributional-jensen-results}

Table~\ref{tab:jensen-gap} shows the CAFE distributional Jensen Gap. All gaps
are positive and all \(95\%\) bootstrap confidence intervals lie strictly above
zero. Under the CAFE criterion, all architectures therefore expose
antifragility-compatible stress geometry.

\begin{table}[t]
\centering
\caption{Distributional Jensen Gap by architecture. Positive gaps indicate
convex-expansive observed stress deformation relative to the expected stress
distribution.}
\label{tab:jensen-gap}
\begin{tabular}{lrrrr}
\hline
Architecture & Expected disp. & Observed disp. & Jensen Gap & 95\% CI \\
\hline
A0 Flat & 0.1861 & 0.2648 & 0.0787 & [0.0661, 0.0935] \\
A1 Hierarchical & 0.1861 & 0.2728 & 0.0867 & [0.0719, 0.1014] \\
A2 Debate & 0.1861 & 0.2641 & 0.0780 & [0.0647, 0.0927] \\
A3 Meta-Adaptive & 0.1866 & 0.2675 & 0.0809 & [0.0667, 0.0956] \\
A4 Ensemble & 0.1861 & 0.2836 & 0.0975 & [0.0822, 0.1146] \\
\hline
\end{tabular}
\end{table}

\begin{figure}[t]
    \centering
    \includegraphics[width=0.92\linewidth]{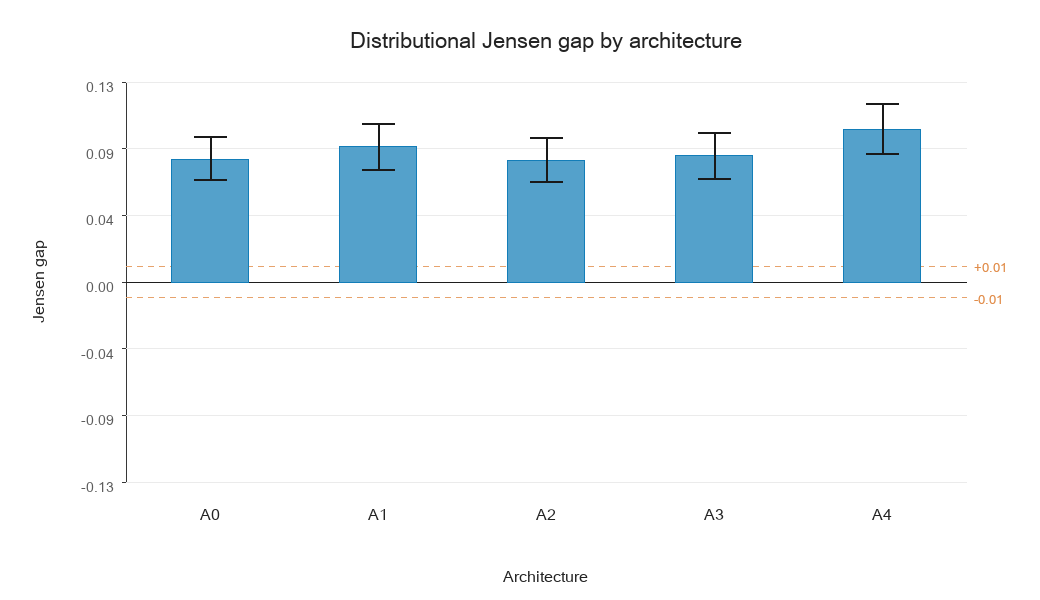}
    \caption{Distributional Jensen Gap by architecture. Error bars denote
    bootstrap \(95\%\) confidence intervals, and dashed lines mark the
    resilience tolerance.}
    \label{fig:jensen-gap-by-arch}
\end{figure}

Figure~\ref{fig:jensen-gap-by-arch} visualizes the same result. The dashed
lines mark the resilience band. Every confidence interval lies above the upper
resilience threshold, so the conclusion does not depend on small numerical
fluctuations around zero.

The ranking by antifragility-compatible opportunity is
\[
    \mathrm{A4} > \mathrm{A1} > \mathrm{A3} > \mathrm{A0} > \mathrm{A2}.
\]
This ranking should not be read as a ranking of immediate output quality. A4
has the largest gap and the largest quality drop. The correct interpretation is
that A4 exposes the largest amount of structured stress variation, and thus the
largest opportunity for antifragile adaptation.

\subsection{Stress-Deformation Patterns}
\label{subsec:stress-deformation-patterns}

Table~\ref{tab:std} reports the increase in observed standard deviation
relative to the designed stress distribution. These values explain where the
positive Jensen Gap comes from.

\begin{table}[t]
\centering
\caption{Increase in observed stress standard deviation relative to the
designed stress distribution.}
\label{tab:std}
\begin{tabular}{lrrrr}
\hline
Architecture & Conflict & Load & Ambiguity & Drift \\
\hline
A0 Flat & 0.0474 & 0.0178 & 0.0602 & 0.0457 \\
A1 Hierarchical & 0.0568 & 0.0169 & 0.0614 & 0.0513 \\
A2 Debate & 0.0597 & 0.0238 & 0.0279 & 0.0538 \\
A3 Meta-Adaptive & 0.0372 & 0.0284 & 0.0497 & 0.0576 \\
A4 Ensemble & 0.0563 & 0.0372 & 0.0607 & 0.0517 \\
\hline
\end{tabular}
\end{table}

The dominant expansion aligns with architectural structure. A2 Debate expands
most strongly along conflict, consistent with adversarial disagreement. A3
Meta-Adaptive expands most strongly along drift, consistent with its
coherence-and-drift monitoring role. A0, A1, and A4 expand most strongly along
ambiguity, with A4 also showing broad expansion across conflict, ambiguity, and
drift. These patterns suggest that different coordination strategies expose
different learnable stress modes.

\begin{figure}[t]
    \centering
    \begin{minipage}{0.32\linewidth}
        \centering
        \includegraphics[width=\linewidth]{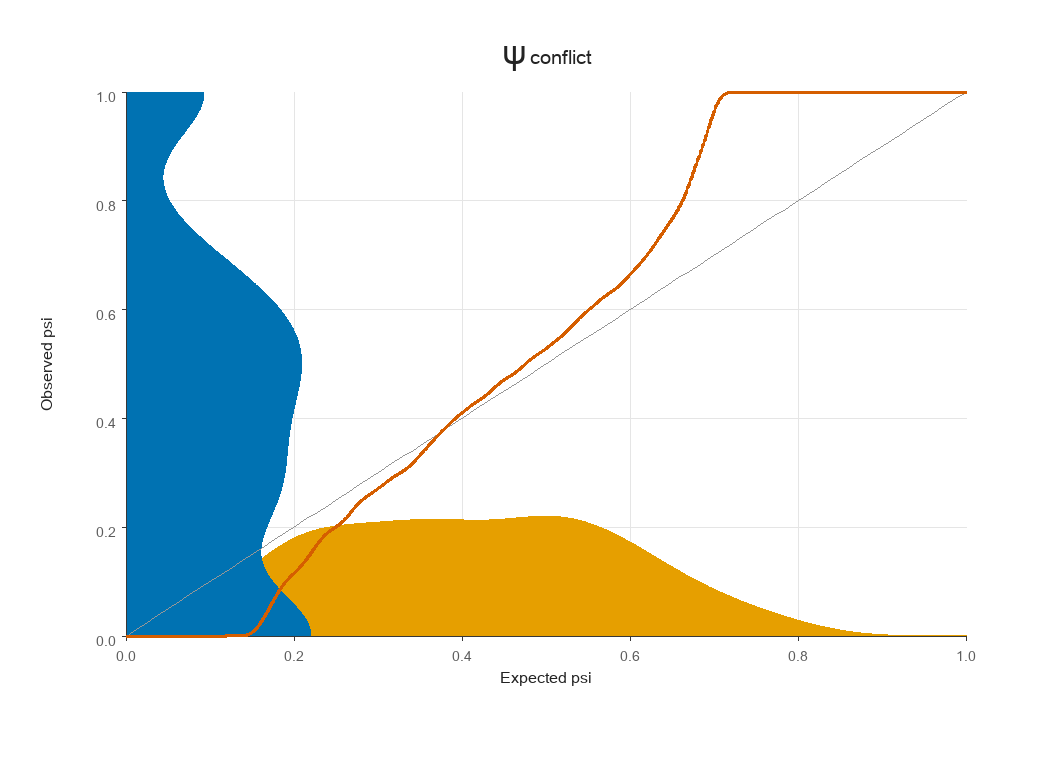}\\
        \small A2 conflict
    \end{minipage}
    \hfill
    \begin{minipage}{0.32\linewidth}
        \centering
        \includegraphics[width=\linewidth]{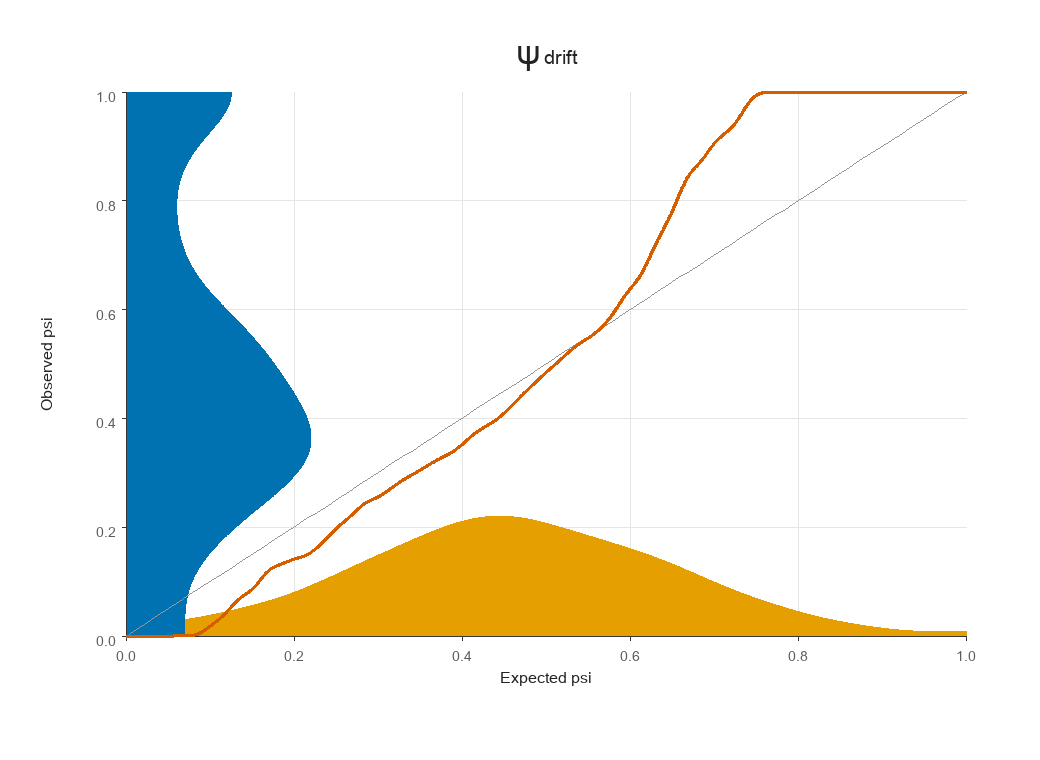}\\
        \small A3 drift
    \end{minipage}
    \hfill
    \begin{minipage}{0.32\linewidth}
        \centering
        \includegraphics[width=\linewidth]{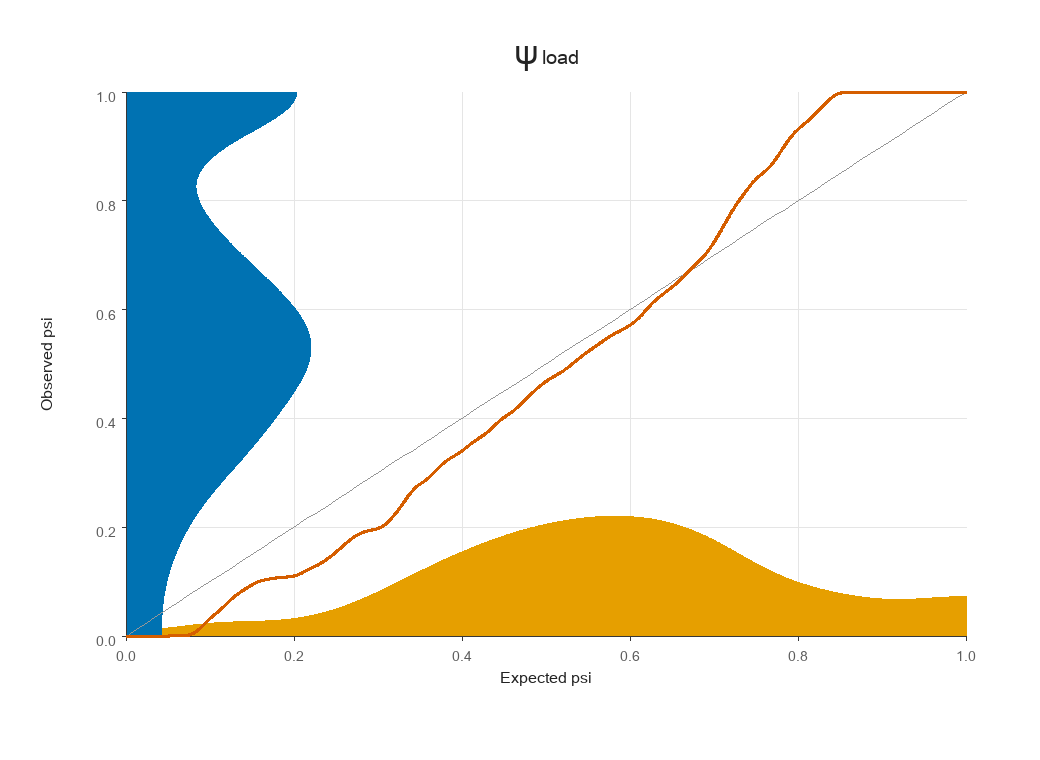}\\
        \small A4 load
    \end{minipage}
    \caption{Representative expected-to-observed marginal deformations.
    These examples show how different coordination mechanisms expose different
    stress modes: adversarial debate expands conflict, meta-adaptive control
    expands drift, and ensemble consensus expands load.}
    \label{fig:selected-marginals}
\end{figure}

Figure~\ref{fig:selected-marginals} shows representative marginal deformation
plots. We keep the full set of marginal diagnostics in
Appendix~\ref{app:marginal-diagnostics}; the main text reports only the
examples that best explain the architecture-specific patterns in
Table~\ref{tab:std}.

\subsection{Response-Surface Fit Diagnostics}
\label{subsec:fit-diagnostics}

Table~\ref{tab:fit-mean} summarizes response-surface fit quality averaged over
the four judge dimensions. The fits are moderate but nontrivial, supporting the
use of the response model for inverse stress reconstruction while motivating
bootstrap uncertainty estimates.

\begin{table}[t]
\centering
\caption{Mean response-surface fit diagnostics across judge dimensions.}
\label{tab:fit-mean}
\begin{tabular}{lrrr}
\hline
Architecture & Mean \(R^2\) & Mean RMSE & Mean MAE \\
\hline
A0 Flat & 0.4888 & 0.1044 & 0.0783 \\
A1 Hierarchical & 0.4549 & 0.1164 & 0.0886 \\
A2 Debate & 0.4794 & 0.1126 & 0.0847 \\
A3 Meta-Adaptive & 0.4672 & 0.1149 & 0.0876 \\
A4 Ensemble & 0.4488 & 0.1293 & 0.0940 \\
\hline
\end{tabular}
\end{table}

Structural drift inverse is the most predictable judge dimension, with
\(R^2\) values between \(0.669\) and \(0.717\). Novel inference rate is also
moderately explained, with \(R^2\) between \(0.449\) and \(0.574\).
Contradiction resolution is noisier, with \(R^2\) between \(0.241\) and
\(0.346\), reflecting the difficulty of modeling fine-grained semantic
reconciliation.

\subsection{Summary}
\label{subsec:results-summary}

The results support the intended interpretation of CAFE. The framework does
not certify that a multi-agent architecture is already antifragile in the sense
of improving output quality under stress. Instead, it detects whether the
architecture enters a statistically antifragility-compatible regime. In all
five architectures, stress lowers average quality but expands the observed
stress distribution. This means the systems expose structured variation under
adversity. Such variation is the raw material for future antifragile learning:
adaptive routing, memory updates, architecture selection, or feedback-driven
coordination could use this signal to improve from adversity rather than merely
endure it.

\section{Discussion}
\label{sec:discussion}

The results support a deliberately narrow interpretation of CAFE. The evaluated
architectures do not improve their average judged quality under semantic
stress. In fact, all architectures exhibit substantial quality degradation
under perturbation. However, this degradation is accompanied by a positive
distributional Jensen Gap in every case. The key implication is therefore not
that the current architectures are already antifragile in an operational sense,
but that they enter regimes in which stress produces structured,
convex-expansive variation in the observed effective stress distribution. CAFE is a
detector of this opportunity.

\subsection{Quality Degradation and Antifragility Opportunity}
\label{subsec:quality-vs-opportunity}

A central risk in studying antifragility is to conflate two different claims.
The first claim is immediate performance improvement: a system receives a
harder input and produces a better output. Our experiments do not support that
claim. Mean judge quality decreases by roughly one third across all
architectures. The second claim is statistical opportunity: a system's response
to stress becomes more informative, more differentiated, or more structured in
a way that a future adaptive mechanism could exploit. The positive Jensen Gaps
support this second claim.

This distinction is important for the interpretation of CAFE. A positive gap is
not a certificate of final task success. It indicates that the observed stress
distribution has expanded relative to the expected stress distribution under a
convex potential. In practical terms, the architecture exposes more variation
about how it experiences contradiction, overload, ambiguity, and drift. Such
variation is useful only if there is a mechanism that can learn from it:
adaptive routing, memory updates, stress-aware model selection, prompt repair,
or architecture reconfiguration. Without such a mechanism, the system may
remain brittle despite having an antifragility-compatible signal.

\subsection{Architectural Effects}
\label{subsec:architectural-effects}

The architecture ranking induced by CAFE differs from a ranking based on
immediate quality. A4 Ensemble has the largest quality drop and also the
largest Jensen Gap. This makes A4 a useful example of the intended reading of
the metric: the ensemble does not solve the stressed prompts better in the
short term, but it exposes the richest observed stress geometry. The diversity
of independent analysts appears to create broader variation across stress
dimensions, especially load, ambiguity, and drift.

A1 Hierarchical has the second-largest gap, suggesting that decomposition into
risk-specific specialists also increases stress differentiation. This is
consistent with a division-of-labor interpretation: different specialists may
react differently to semantic overload, ambiguous references, or temporal
inconsistency, even if the final synthesized answer still loses quality under
stress.

A2 Debate expands most strongly along conflict, which matches its adversarial
structure. The debate architecture is designed to surface disagreement, so it
is expected to be sensitive to contradictory evidence. However, A2 has the
smallest global Jensen Gap among the five architectures. This suggests that
adversarial debate may expose one stress mode sharply without producing the
broadest distributional expansion across the full stress space.

A3 Meta-Adaptive expands most strongly along drift. This is consistent with a
meta-controller that reacts to coherence and temporal inconsistency. The result
is promising but not conclusive: the current meta-adaptive architecture exposes
drift-related variation, but the experiment does not yet close the loop by
allowing the controller to learn from repeated exposure.

\subsection{What the Jensen Gap Measures}
\label{subsec:jensen-gap-interpretation}

The distributional Jensen Gap should be read as a geometric statistic, not as a
generic distance between distributions. A generic divergence can say that the
expected and observed distributions differ, but it does not specify whether the
difference corresponds to expansion or compression under a convex stress
potential. CAFE focuses on this directionality. Positive values indicate that
the observed effective stress distribution is more dispersed than the designed
stress distribution; negative values would indicate compression; values close
to zero indicate resilience or preservation of the designed stress geometry.

This makes the choice of potential \(\phi\) conceptually important. In this
paper we use \(\phi(\mathbf{x})=\lVert\mathbf{x}\rVert_2^2\), which treats all
stress dimensions symmetrically and measures total dispersion. This is a
conservative first choice because it avoids hand-tuning the metric to favor one
architecture. In future work, domain-specific potentials could assign greater
weight to particular stresses, such as contradiction in factual verification
tasks or temporal drift in longitudinal decision support.

\subsection{Limitations}
\label{subsec:discussion-limitations}

The first limitation is that observed stress is reconstructed rather than
directly measured. The inverse problem is constrained by four judge signals,
regularization, and the fitted polynomial response model. This makes the
observed distribution an effective latent quantity, not a ground-truth physical
measurement. The fit diagnostics show moderate explanatory power, which is
enough for a first controlled study but not enough to treat individual
reconstructed stress vectors as exact. For this reason, CAFE should be used as
a distributional diagnostic rather than as a per-sample stress oracle.

The second limitation is that the current experiment is single-domain. Banking
risk is a useful stress-test domain because it naturally combines conflicting
evidence, temporal drift, overloaded context, and ambiguity. Nevertheless, the
stress geometry may differ in medical reasoning, legal analysis, software
engineering, scientific question answering, or open-ended planning. A stronger
claim would require repeating the analysis across domains with different
stress semantics.

The third limitation is that CAFE identifies opportunity but does not implement
the adaptation loop. The present study stops after detecting convex-expansive
stress deformation. It does not update memories, retrain routers, change
architecture selection policies, or optimize prompts based on the detected
stress modes. Therefore, the paper should not claim that the evaluated
architectures already improve from adversity. It should claim that CAFE
identifies when such improvement may be learnable.

The fourth limitation concerns the judge. The response signals are produced by
an automatic evaluator, and the reconstruction depends on their stability.
Although the judge is deterministic and multi-dimensional, it may still encode
rubric biases. Human calibration, judge ensembles, or cross-judge sensitivity
analysis would strengthen the empirical claim.

\subsection{Implications}
\label{subsec:discussion-implications}

The main implication is that robustness and antifragility should be separated
in the evaluation of multi-agent LLM systems. Robustness asks whether quality
is preserved under perturbation. CAFE asks whether stress produces a structured
signal that could support future improvement. These are related but distinct
properties. A system can be non-robust in immediate quality while still exposing
an antifragility-compatible stress geometry.

This perspective suggests a concrete research program. First, use CAFE to
detect which architectures and stress dimensions produce positive
distributional expansion. Second, attach adaptive mechanisms to those stress
modes: route high-conflict cases to debate, route high-drift cases to temporal
verification, or update memory when repeated overload patterns appear. Third,
measure whether repeated use of these mechanisms turns antifragility-compatible
opportunity into actual performance gains over time. Under this view, CAFE is
not the final antifragile system. It is the measurement layer needed to decide
where antifragile learning should be applied.

\section{Conclusions and Future Work}
\label{sec:conclusion-future-work}

We introduced CAFE, a statistical framework for identifying
antifragility-compatible regimes in multi-agent LLM architectures. CAFE
separates immediate performance from antifragility opportunity. In our
experiments, semantic stress reduces average judge quality across all
architectures, so the systems should not be described as already antifragile in
the strong operational sense. At the same time, every architecture exhibits a
positive distributional Jensen Gap: the observed effective stress distribution
expands relative to the expected stress distribution under a convex potential.
This suggests that the stressed systems expose structured variation that could
be exploited by future adaptation.

The contribution of CAFE is therefore methodological. It provides a way to ask
whether a given stress scenario is statistically antifragility-compatible before
building the adaptive mechanism that would learn from it. This is a useful
separation: without a measurement layer, antifragility learning risks optimizing
against noisy degradation rather than against structured stress signals.

\subsection{Antifragility Learning}
\label{subsec:future-antifragility-learning}

The first future direction is to design explicit antifragility learning
methods. In the current paper, CAFE detects where useful stress variation
appears; it does not yet convert that variation into improved behavior. A next
step is to use the recovered stress modes as training or control signals. For
example, high-conflict observations could trigger debate-based routing,
high-drift observations could trigger temporal verification, and high-load
observations could trigger decomposition or summarization before final
synthesis.

More generally, antifragility learning should optimize architectures to
benefit from repeated exposure to stress. This could be implemented through
stress-aware routers, memory systems that store failure modes, curriculum
policies that sample perturbations with positive CAFE gaps, or meta-controllers
that select coordination strategies based on the inferred stress profile. The
key open question is whether positive distributional Jensen Gaps can be turned
into later gains in quality, calibration, factuality, or robustness.

\subsection{Evolution of Adapted Multi-Agent Systems}
\label{subsec:future-agent-evolution}

The second direction is to study how the antifragile capacity of a multi-agent
system changes after it has been adapted with antifragility learning. The
present experiments evaluate each architecture at a single point: prompts are
generated, responses are judged, observed stress is reconstructed, and the
Jensen Gap is estimated. In an adaptive MAS, this measurement should become a
trajectory. After each adaptation step, CAFE can estimate
\(\mathcal{G}_{a,t}\) across time windows and quantify whether the system's
antifragility-compatible regime strengthens, weakens, shifts across stress
dimensions, or disappears.

This joint temporal and architectural view is important because adaptation can
change both performance and stress geometry. A system may initially show a
positive gap because stress exposes many unresolved failure modes. After
adaptation, the same gap could decrease because the system has learned to
absorb the stress more smoothly. Alternatively, the gap could remain positive
while quality improves, indicating that the system continues to extract useful
variation from new stressors. The relevant question is therefore not only
whether the MAS improves, but how its antifragile capacity changes: by how much
the gap changes, which stress dimensions drive the change, and whether the
change is stable over repeated stress exposure.

Current multi-agent architectures are usually specified manually: flat
pipelines, hierarchies, debates, meta-controllers, or ensembles. A CAFE-driven
system could instead adapt its topology as its stress geometry evolves. It
could add specialists when a stress dimension repeatedly expands, prune agents
that do not contribute useful stress information, or switch between debate,
hierarchy, and ensemble modes depending on the inferred stress regime.

This opens a broader question: what kinds of multi-agent organization emerge
when the optimization target is not only immediate answer quality, but also
the ability to learn from adversity? Studying this question requires logging
architecture changes, stress distributions, response quality, and CAFE gaps
over long horizons. The resulting systems may not converge to a single fixed
architecture; they may become adaptive populations of agents whose structure
changes with the stress landscape.

\subsection{Forecasting Antifragile Regimes}
\label{subsec:future-forecasting}

The third direction is forecasting. If CAFE can estimate whether a scenario is
currently antifragility-compatible, a natural next question is whether such
regimes can be predicted before full evaluation. Forecasting could use early
judge signals, partial responses, prompt-level stress descriptors, historical
CAFE trajectories, or architecture metadata to predict the future Jensen Gap
and its uncertainty.

Such forecasting would support proactive control. A system could allocate more
compute to prompts likely to produce useful stress variation, preselect
architectures that historically benefit from a given stress profile, or avoid
adaptation when the forecast suggests concave compression and likely fragility.
In deployment, this would turn CAFE from an offline diagnostic into an online
decision layer for stress-aware multi-agent systems.

\subsection{Closing Remarks}
\label{subsec:closing-remarks}

CAFE reframes antifragility evaluation as a distributional measurement problem.
Rather than asking only whether performance survives perturbation, it asks
whether stress creates a structured signal that future systems can learn from.
The present results show that this signal can be detected even when immediate
quality decreases. The next step is to close the loop: build multi-agent
systems that use CAFE signals to adapt, track whether their antifragility
evolves over time, and forecast when adversity is likely to become useful
rather than merely damaging.

\printbibliography
\newpage
\appendix
\section{Marginal Stress Deformation Diagnostics}
\label{app:marginal-diagnostics}

\begin{figure}[h]
    \centering
    \begin{minipage}{0.48\linewidth}
        \centering
        \includegraphics[width=\linewidth]{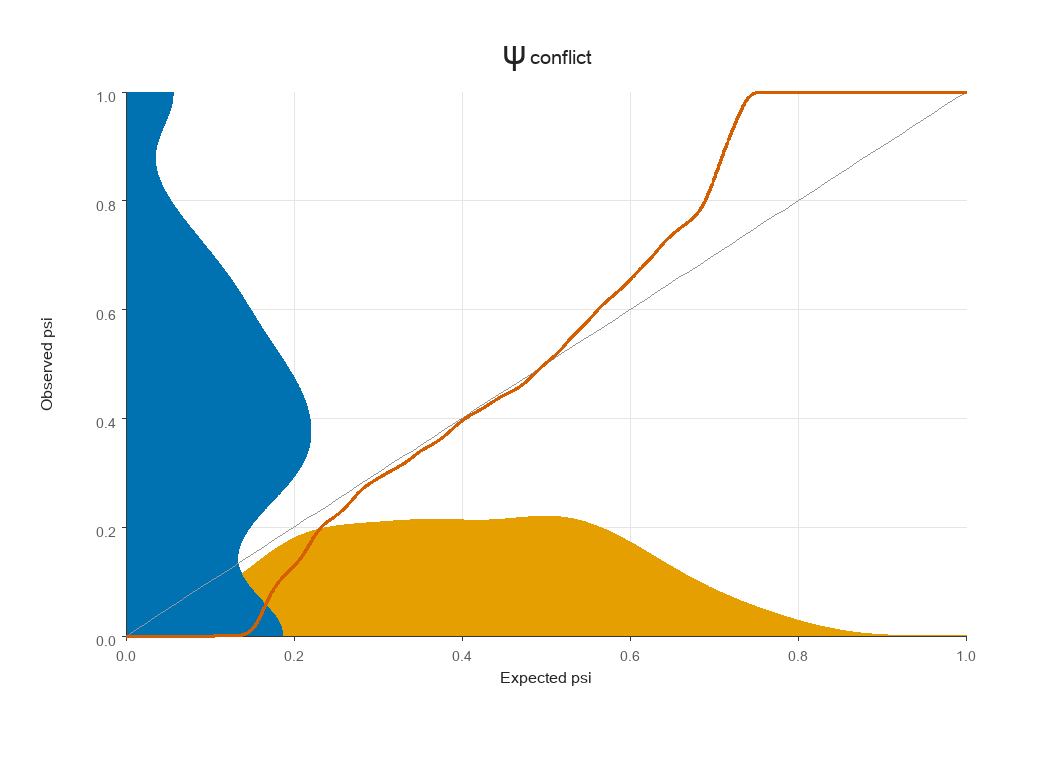}\\
        \small A0 conflict
    \end{minipage}
    \hfill
    \begin{minipage}{0.48\linewidth}
        \centering
        \includegraphics[width=\linewidth]{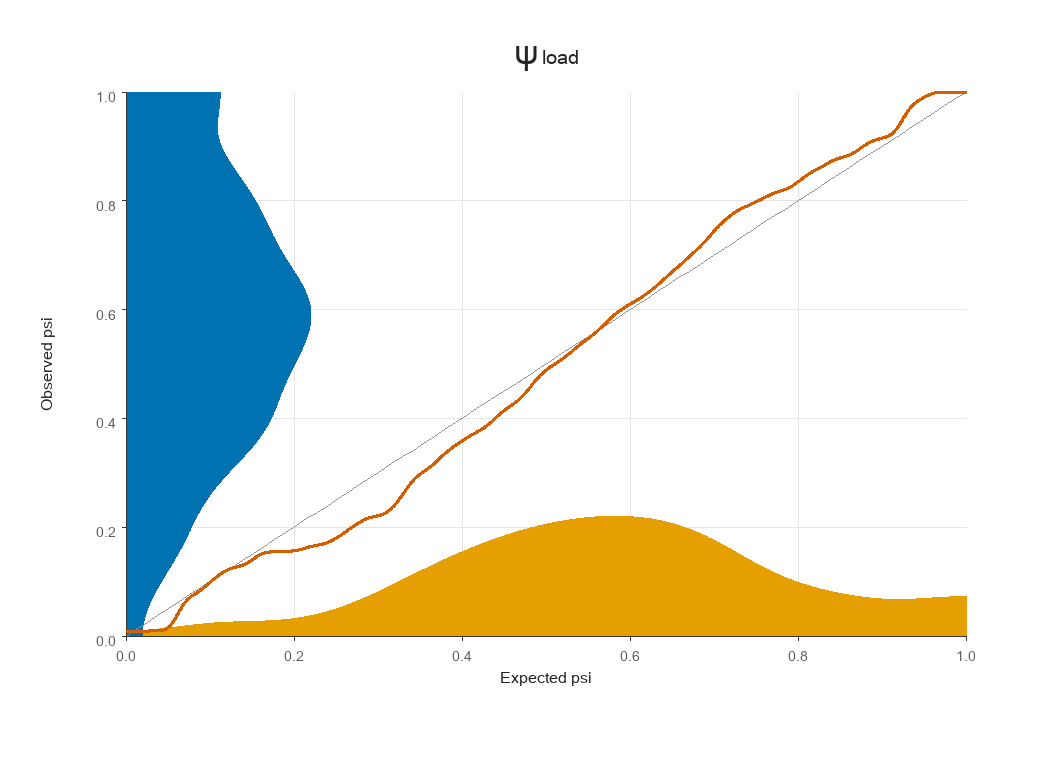}\\
        \small A0 load
    \end{minipage}

    \vspace{0.8em}

    \begin{minipage}{0.48\linewidth}
        \centering
        \includegraphics[width=\linewidth]{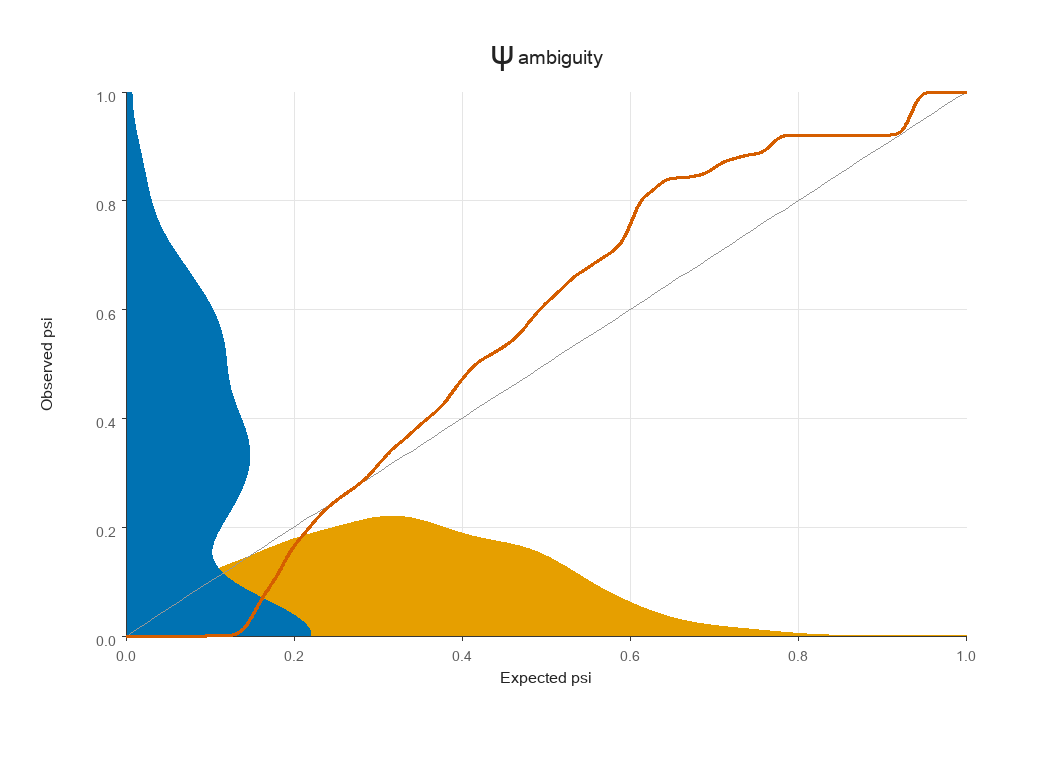}\\
        \small A0 ambiguity
    \end{minipage}
    \hfill
    \begin{minipage}{0.48\linewidth}
        \centering
        \includegraphics[width=\linewidth]{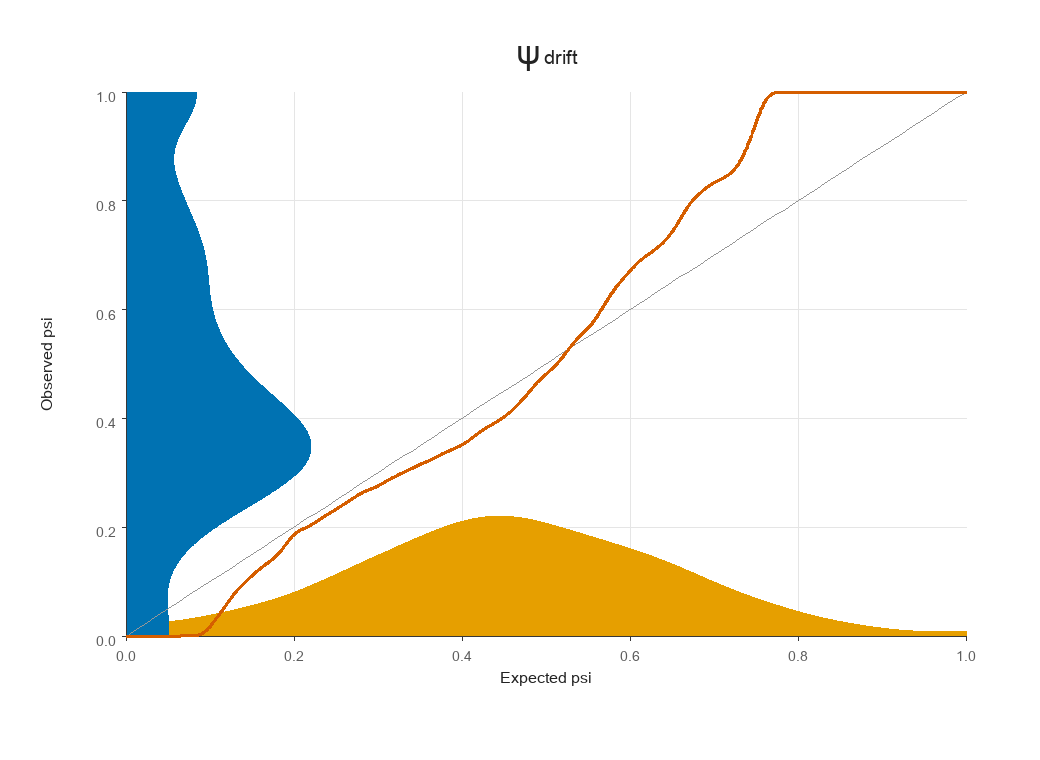}\\
        \small A0 drift
    \end{minipage}
    \caption{Marginal stress deformation diagnostics for A0 Flat.}
    \label{fig:appendix-a0-marginals}
\end{figure}

\begin{figure}[h]
    \centering
    \begin{minipage}{0.48\linewidth}
        \centering
        \includegraphics[width=\linewidth]{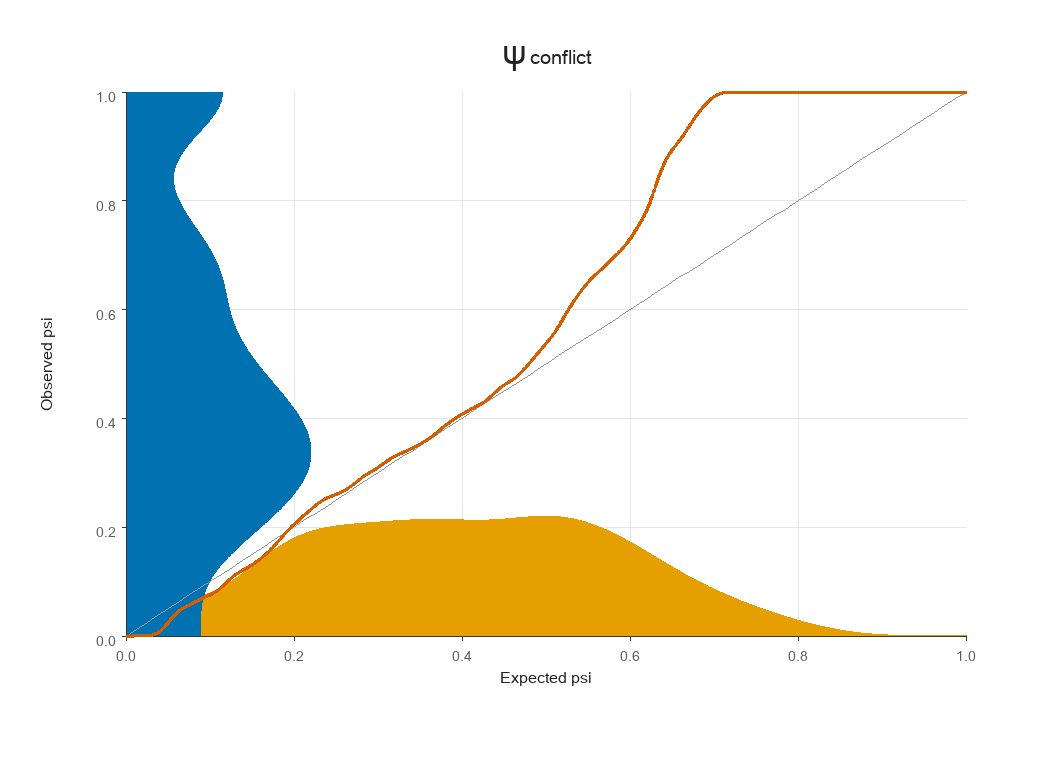}\\
        \small A1 conflict
    \end{minipage}
    \hfill
    \begin{minipage}{0.48\linewidth}
        \centering
        \includegraphics[width=\linewidth]{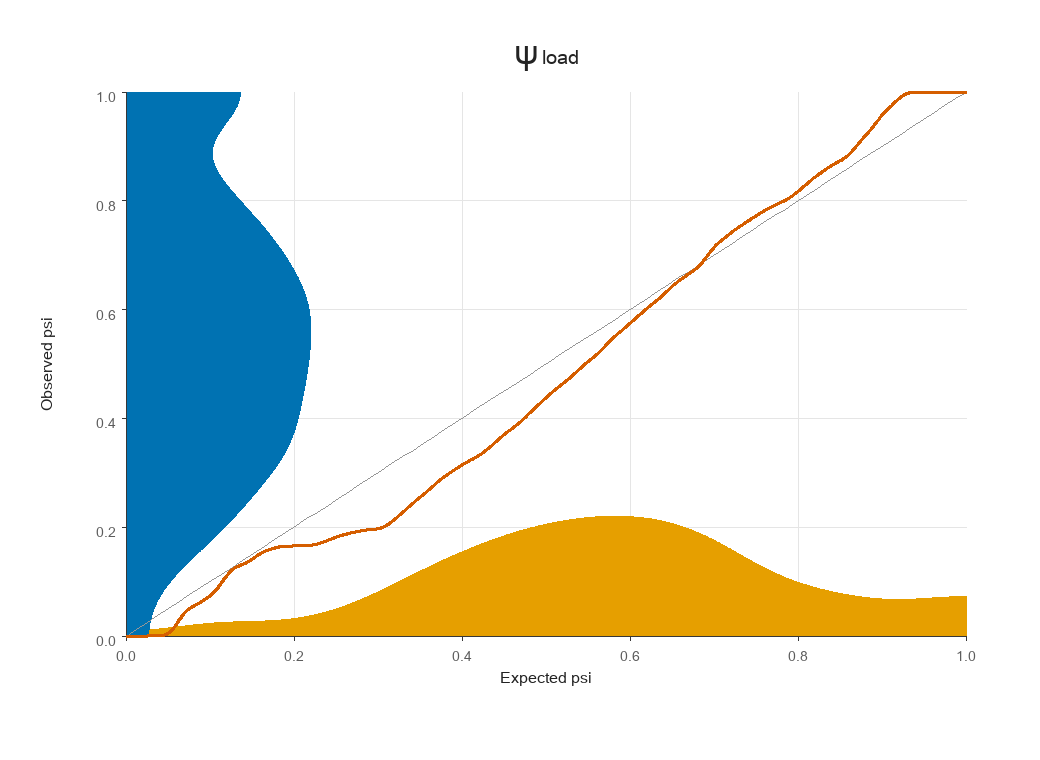}\\
        \small A1 load
    \end{minipage}

    \vspace{0.8em}

    \begin{minipage}{0.48\linewidth}
        \centering
        \includegraphics[width=\linewidth]{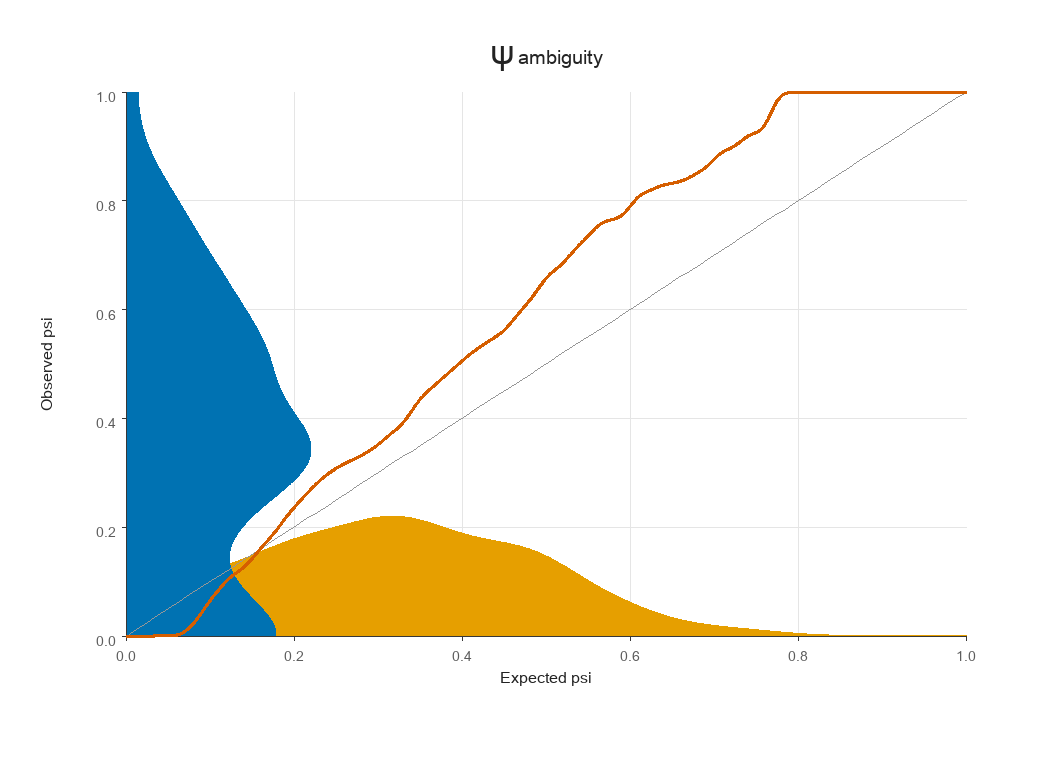}\\
        \small A1 ambiguity
    \end{minipage}
    \hfill
    \begin{minipage}{0.48\linewidth}
        \centering
        \includegraphics[width=\linewidth]{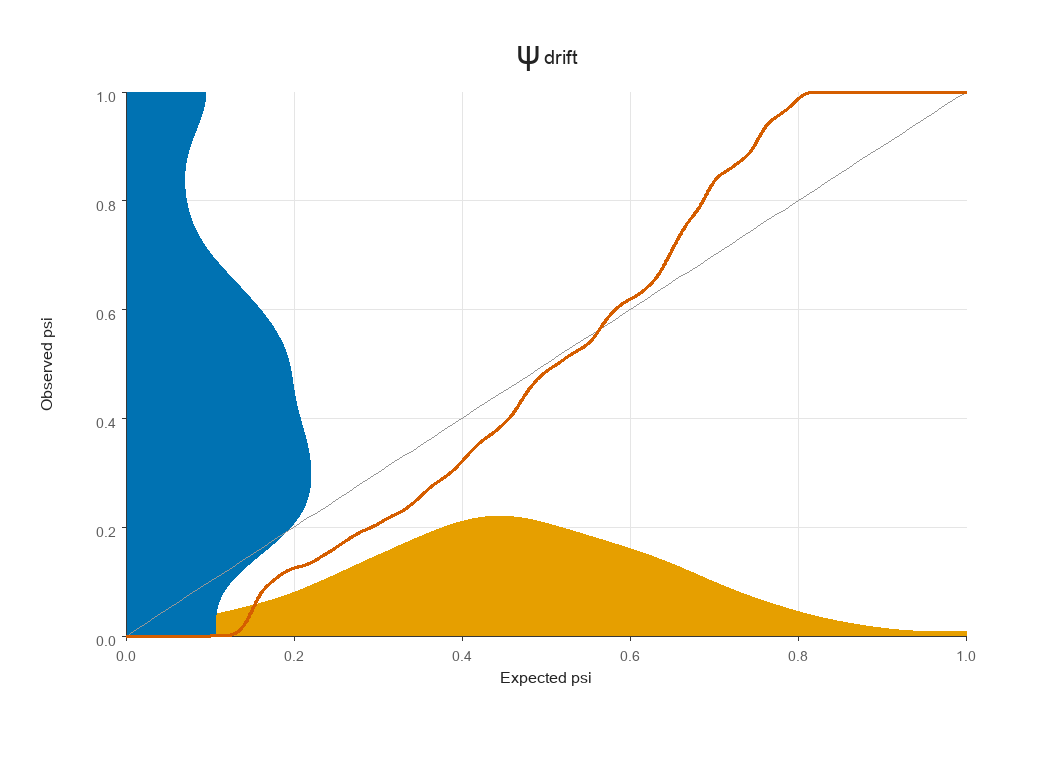}\\
        \small A1 drift
    \end{minipage}
    \caption{Marginal stress deformation diagnostics for A1 Hierarchical.}
    \label{fig:appendix-a1-marginals}
\end{figure}

\begin{figure}[h]
    \centering
    \begin{minipage}{0.48\linewidth}
        \centering
        \includegraphics[width=\linewidth]{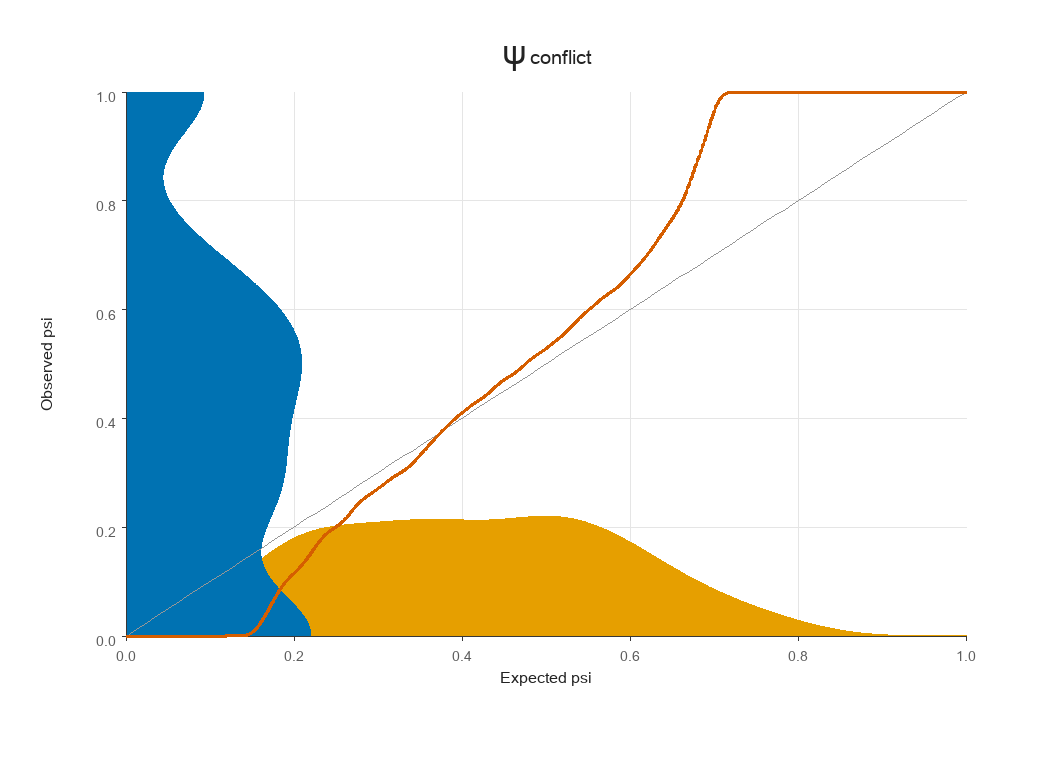}\\
        \small A2 conflict
    \end{minipage}
    \hfill
    \begin{minipage}{0.48\linewidth}
        \centering
        \includegraphics[width=\linewidth]{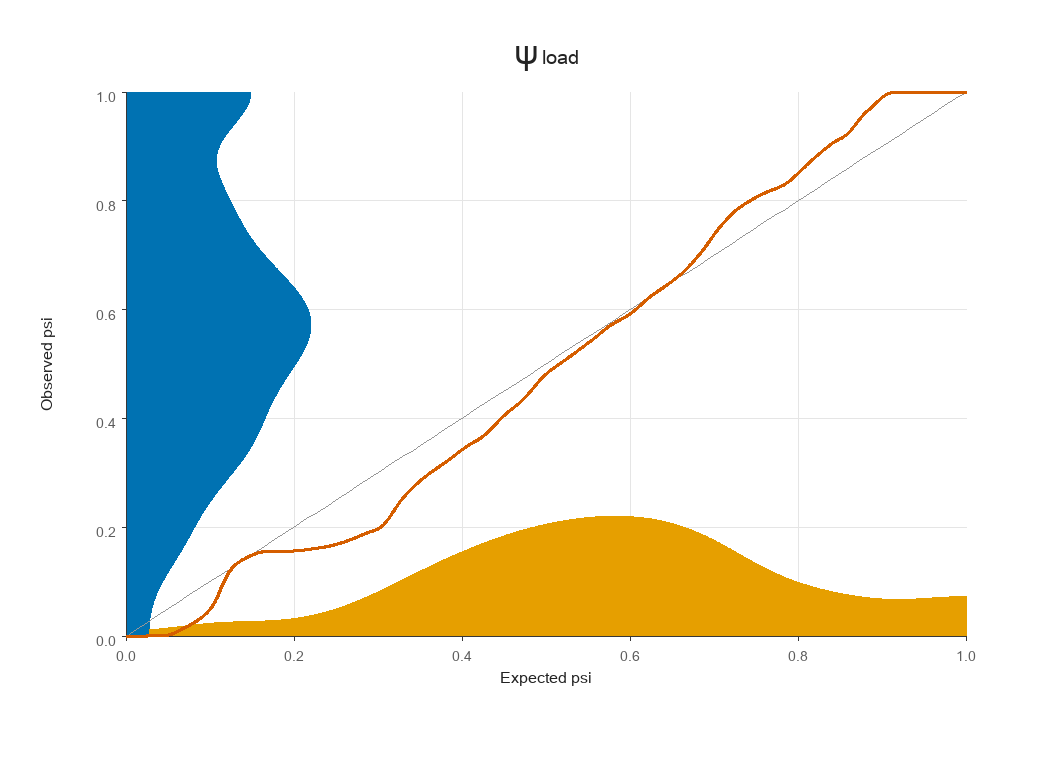}\\
        \small A2 load
    \end{minipage}

    \vspace{0.8em}

    \begin{minipage}{0.48\linewidth}
        \centering
        \includegraphics[width=\linewidth]{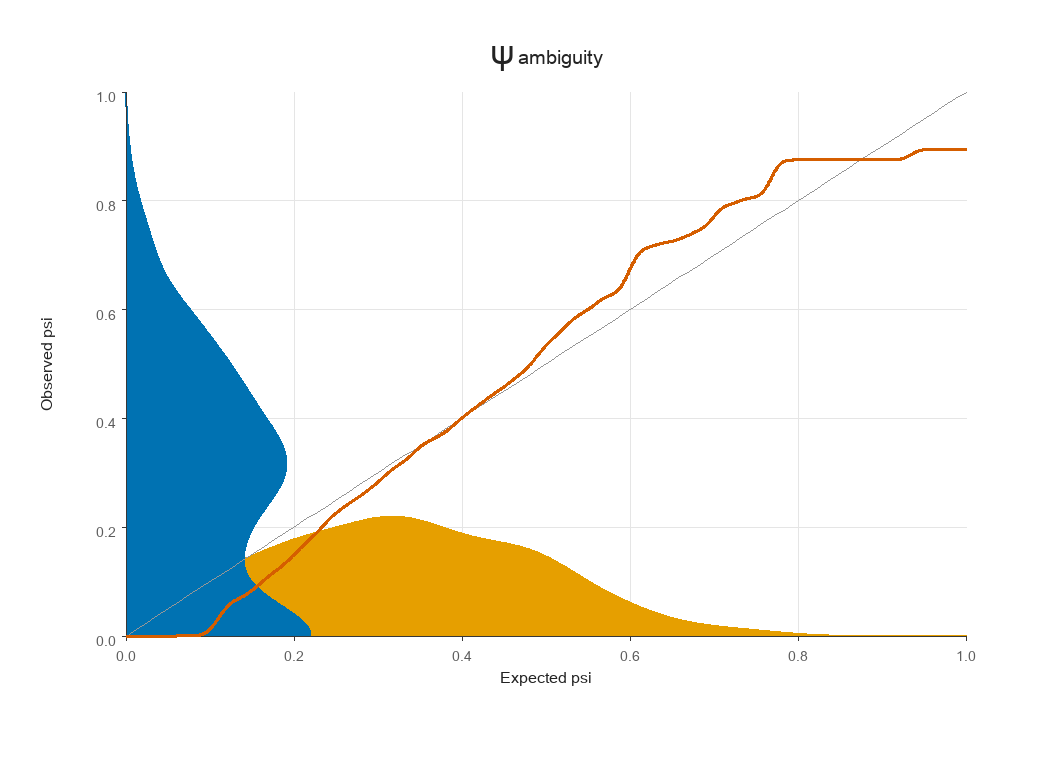}\\
        \small A2 ambiguity
    \end{minipage}
    \hfill
    \begin{minipage}{0.48\linewidth}
        \centering
        \includegraphics[width=\linewidth]{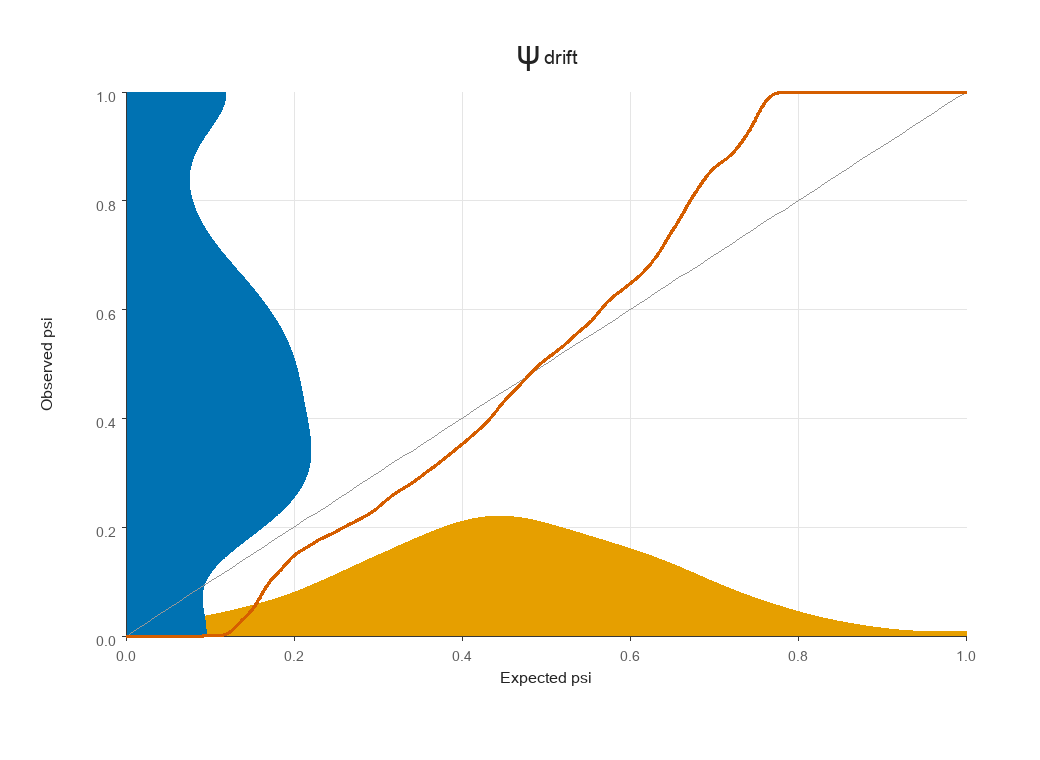}\\
        \small A2 drift
    \end{minipage}
    \caption{Marginal stress deformation diagnostics for A2 Adversarial Debate.}
    \label{fig:appendix-a2-marginals}
\end{figure}

\begin{figure}[h]
    \centering
    \begin{minipage}{0.48\linewidth}
        \centering
        \includegraphics[width=\linewidth]{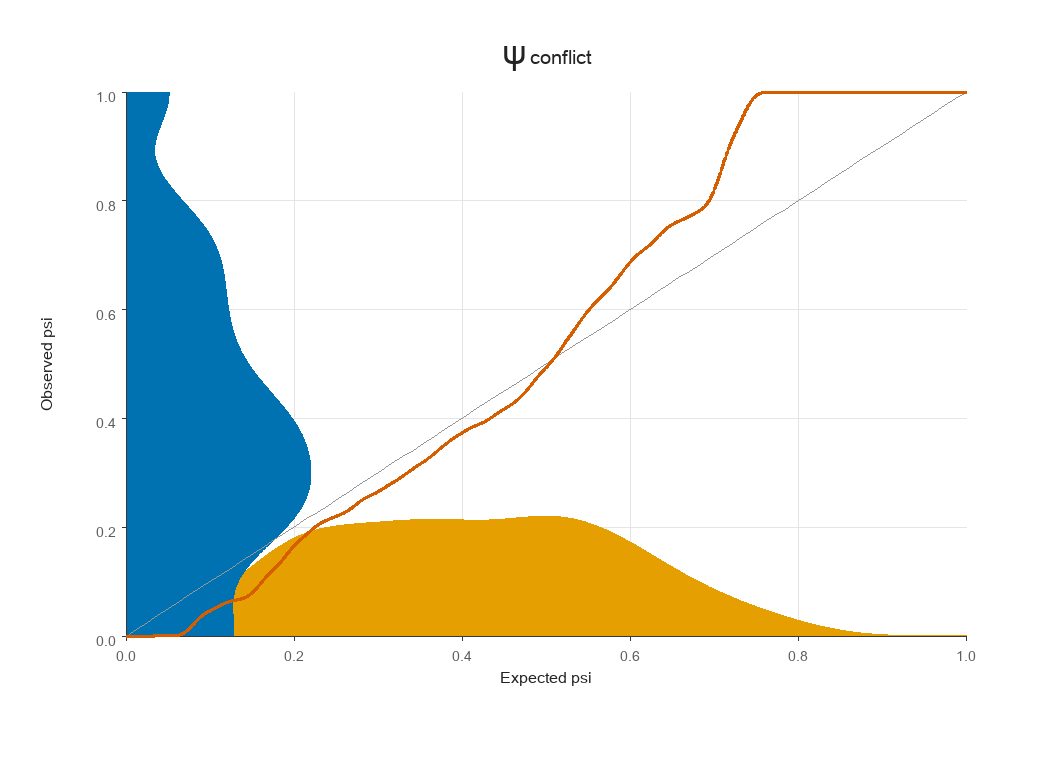}\\
        \small A3 conflict
    \end{minipage}
    \hfill
    \begin{minipage}{0.48\linewidth}
        \centering
        \includegraphics[width=\linewidth]{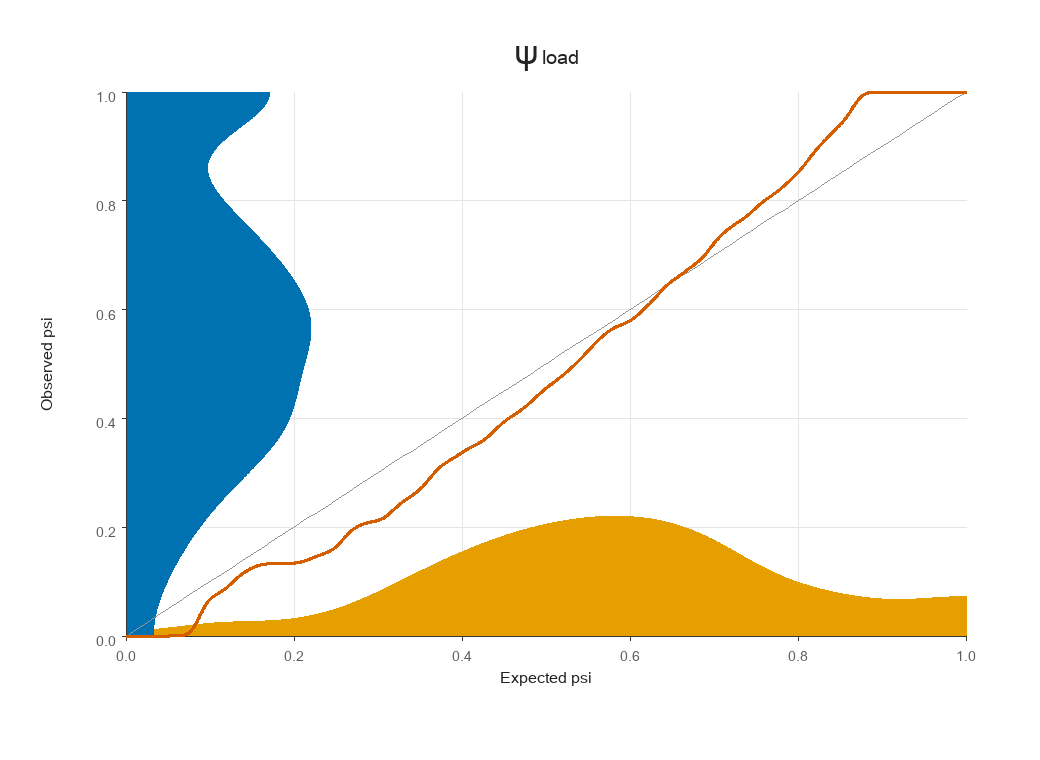}\\
        \small A3 load
    \end{minipage}

    \vspace{0.8em}

    \begin{minipage}{0.48\linewidth}
        \centering
        \includegraphics[width=\linewidth]{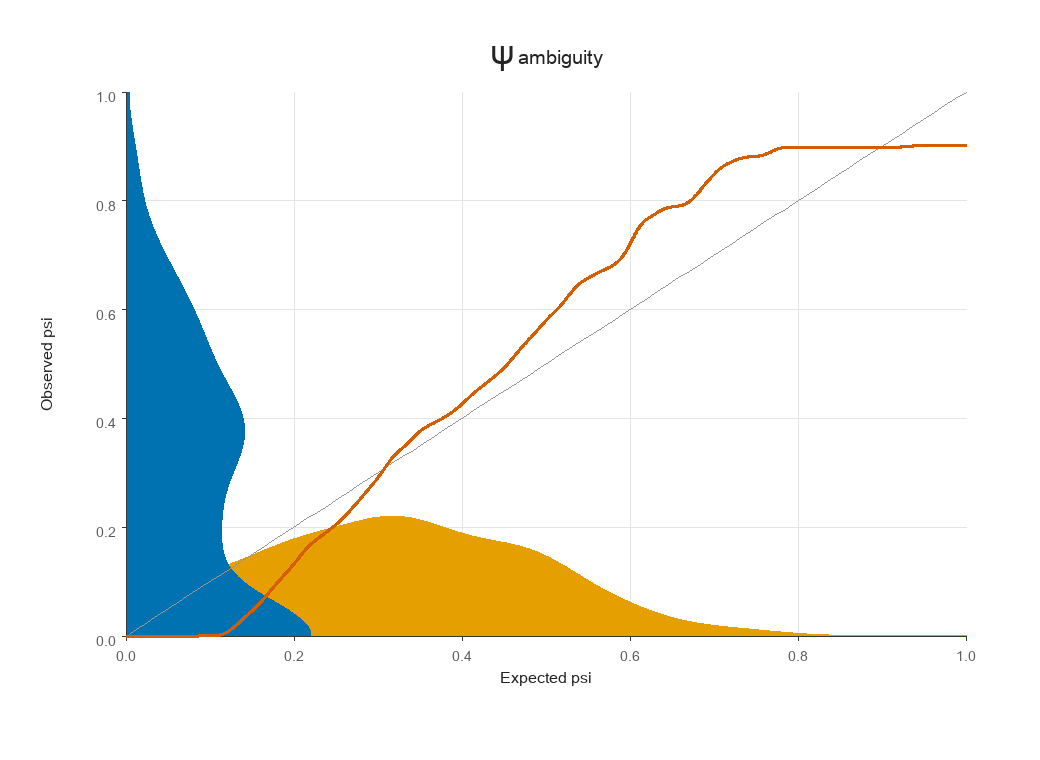}\\
        \small A3 ambiguity
    \end{minipage}
    \hfill
    \begin{minipage}{0.48\linewidth}
        \centering
        \includegraphics[width=\linewidth]{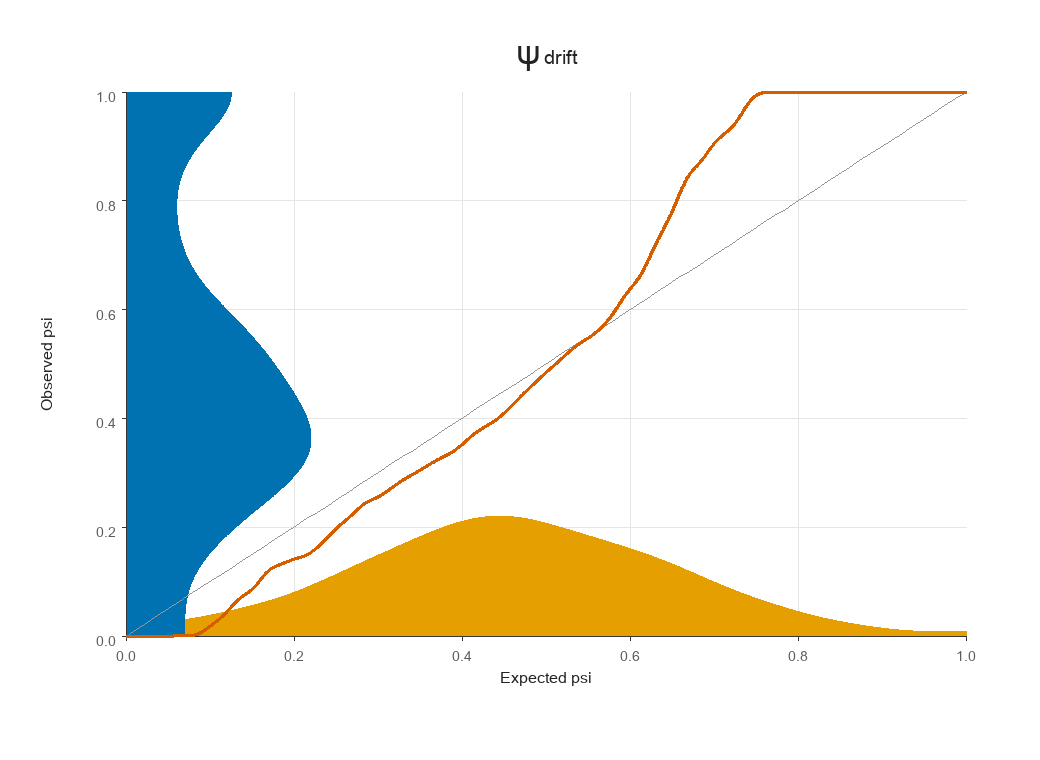}\\
        \small A3 drift
    \end{minipage}
    \caption{Marginal stress deformation diagnostics for A3 Meta-Adaptive.}
    \label{fig:appendix-a3-marginals}
\end{figure}

\begin{figure}[h]
    \centering
    \begin{minipage}{0.48\linewidth}
        \centering
        \includegraphics[width=\linewidth]{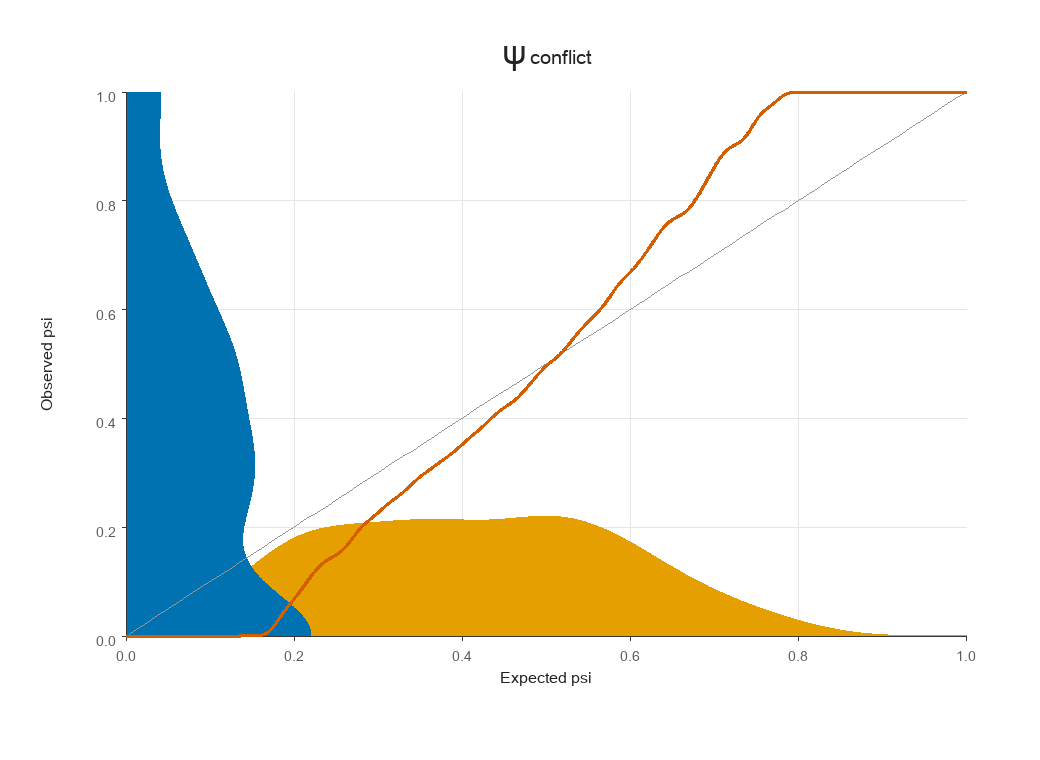}\\
        \small A4 conflict
    \end{minipage}
    \hfill
    \begin{minipage}{0.48\linewidth}
        \centering
        \includegraphics[width=\linewidth]{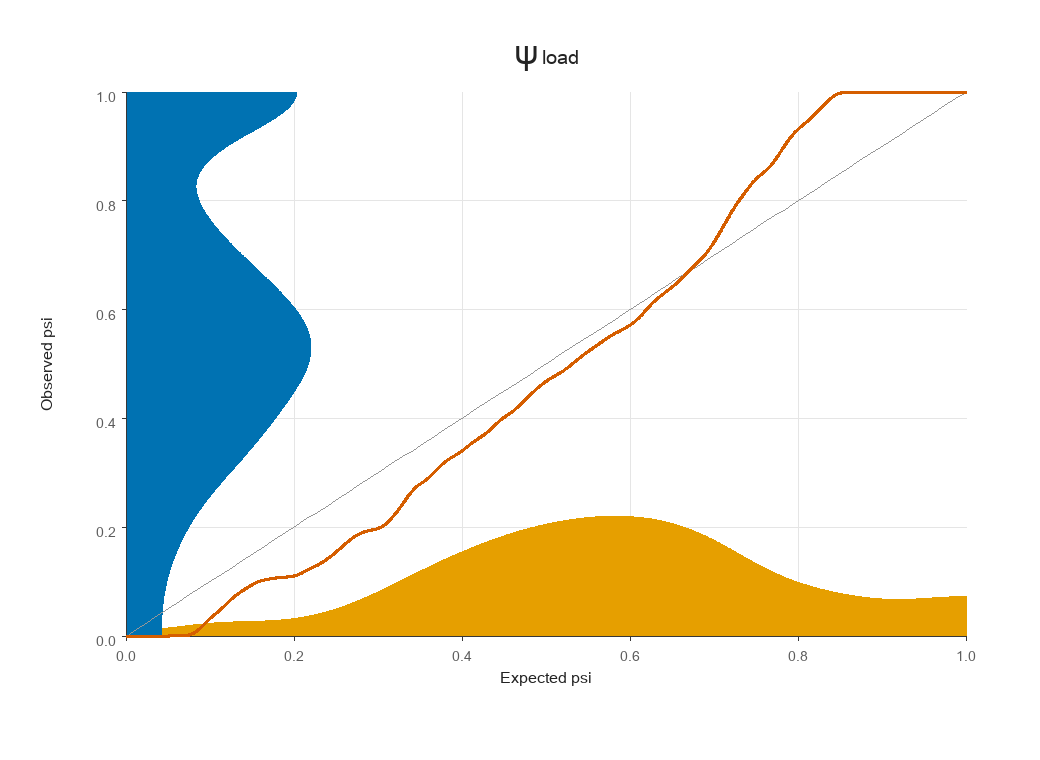}\\
        \small A4 load
    \end{minipage}

    \vspace{0.8em}

    \begin{minipage}{0.48\linewidth}
        \centering
        \includegraphics[width=\linewidth]{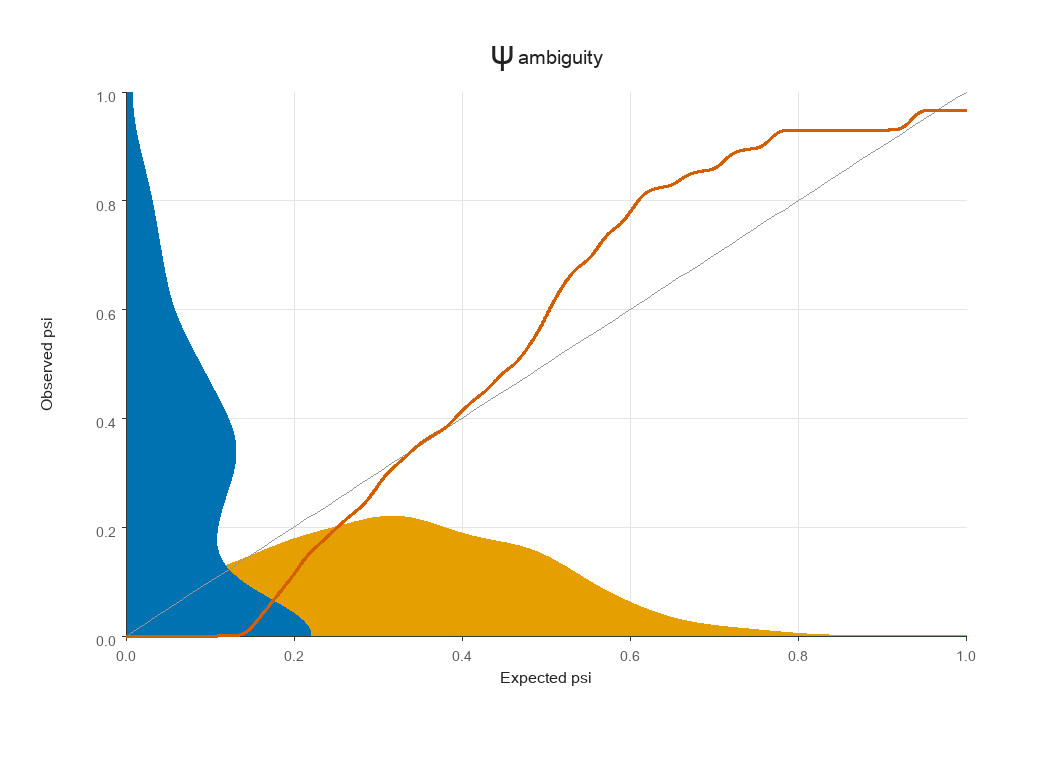}\\
        \small A4 ambiguity
    \end{minipage}
    \hfill
    \begin{minipage}{0.48\linewidth}
        \centering
        \includegraphics[width=\linewidth]{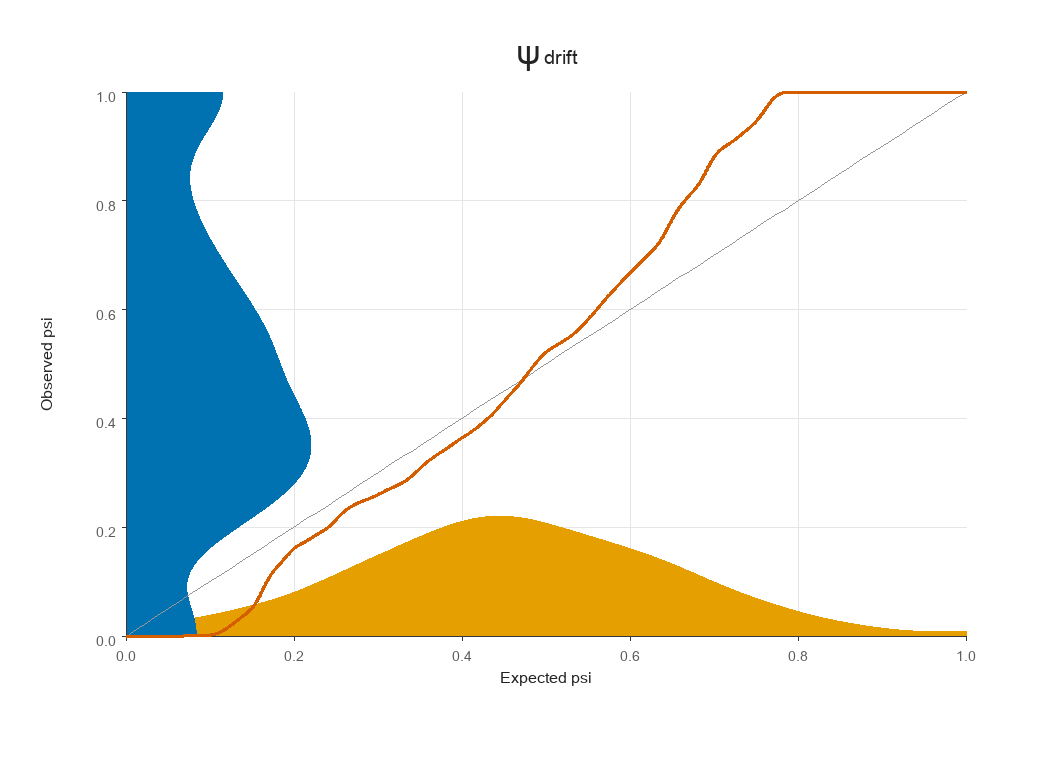}\\
        \small A4 drift
    \end{minipage}
    \caption{Marginal stress deformation diagnostics for A4 Ensemble.}
    \label{fig:appendix-a4-marginals}
\end{figure}

This appendix reports the complete set of marginal expected-to-observed stress
deformation plots. The main results use the aggregate distributional Jensen Gap
and a small number of representative marginals. The figures below are included
to make the architecture-specific deformation patterns auditable. In each
panel, the horizontal axis corresponds to the designed stress intensity
\(\psi_i\), the vertical axis corresponds to the reconstructed observed
effective stress intensity, and the diagonal line denotes no marginal
deformation. Figures~\ref{fig:appendix-a0-marginals}--\ref{fig:appendix-a4-marginals}
report the complete marginal diagnostics for A0 through A4.

\newpage
\section{Additional Statistical Diagnostics}
\label{app:statistical-diagnostics}

This appendix reports response-surface diagnostics at the level of individual
judge signals. The main text reports averages across the four judge
dimensions; Table~\ref{tab:appendix-fit-by-score} shows the full breakdown
used to assess whether the polynomial response model provides enough structure
for inverse stress reconstruction. Table~\ref{tab:appendix-data-diagnostics}
summarizes the number of rows used for each architecture after data
consolidation.

\begin{table}[h]
\centering
\small
\caption{Response-surface fit diagnostics by architecture and judge signal.}
\label{tab:appendix-fit-by-score}
\begin{tabular}{llrrr}
\hline
Architecture & Judge signal & \(R^2\) & RMSE & MAE \\
\hline
A0 Flat & Coherence & 0.403 & 0.100 & 0.079 \\
A0 Flat & Novel inference & 0.574 & 0.077 & 0.053 \\
A0 Flat & Contradiction resolution & 0.271 & 0.147 & 0.116 \\
A0 Flat & Structural preservation & 0.708 & 0.094 & 0.066 \\
\hline
A1 Hierarchical & Coherence & 0.371 & 0.105 & 0.084 \\
A1 Hierarchical & Novel inference & 0.516 & 0.091 & 0.062 \\
A1 Hierarchical & Contradiction resolution & 0.241 & 0.166 & 0.134 \\
A1 Hierarchical & Structural preservation & 0.692 & 0.104 & 0.074 \\
\hline
A2 Debate & Coherence & 0.386 & 0.108 & 0.087 \\
A2 Debate & Novel inference & 0.502 & 0.092 & 0.062 \\
A2 Debate & Contradiction resolution & 0.313 & 0.154 & 0.121 \\
A2 Debate & Structural preservation & 0.717 & 0.097 & 0.068 \\
\hline
A3 Meta-Adaptive & Coherence & 0.385 & 0.108 & 0.087 \\
A3 Meta-Adaptive & Novel inference & 0.497 & 0.090 & 0.061 \\
A3 Meta-Adaptive & Contradiction resolution & 0.285 & 0.160 & 0.129 \\
A3 Meta-Adaptive & Structural preservation & 0.702 & 0.102 & 0.073 \\
\hline
A4 Ensemble & Coherence & 0.331 & 0.128 & 0.096 \\
A4 Ensemble & Novel inference & 0.449 & 0.114 & 0.074 \\
A4 Ensemble & Contradiction resolution & 0.346 & 0.162 & 0.126 \\
A4 Ensemble & Structural preservation & 0.669 & 0.113 & 0.080 \\
\hline
\end{tabular}
\end{table}

Table~\ref{tab:appendix-fit-by-score} shows that the strongest fits occur for
structural preservation, with \(R^2\) between 0.669 and 0.717. Novel inference
is moderately predictable, while contradiction resolution is the noisiest
signal. This pattern is expected: temporal or structural drift leaves
relatively stable traces in the response, whereas contradiction handling
depends on fine-grained semantic reconciliation and therefore has higher
residual variance.

\begin{table}[h]
\centering
\small
\caption{Modeling rows by architecture after data consolidation.}
\label{tab:appendix-data-diagnostics}
\begin{tabular}{lrrr}
\hline
Architecture & Clean rows & Perturbed rows & Modeling rows \\
\hline
A0 Flat & 50 & 500 & 550 \\
A1 Hierarchical & 50 & 500 & 550 \\
A2 Debate & 50 & 500 & 550 \\
A3 Meta-Adaptive & 50 & 498 & 548 \\
A4 Ensemble & 50 & 500 & 550 \\
\hline
\end{tabular}
\end{table}

As shown in Table~\ref{tab:appendix-data-diagnostics}, the slightly smaller A3
sample is due to two missing perturbed evaluations after consolidation. The
CAFE statistics are computed using the available rows for each architecture,
and uncertainty is estimated with architecture-specific bootstrap resampling.

\newpage
\section{Prompt and Response Examples}
\label{app:prompt-response-examples}

This appendix provides a concrete example of the synthetic stress generation
and the corresponding multi-agent outputs. The examples are taken from prompt
\(\mathrm{P001}\). Long responses are excerpted for readability, but the
excerpts preserve the risk rating, the handling of stress, and the main
supervisory recommendation pattern. Table~\ref{tab:perturbed-prompt-examples}
reports the perturbed prompt variants, Table~\ref{tab:clean-output-excerpts}
reports clean-prompt output excerpts, and
Table~\ref{tab:perturbed-output-excerpts} reports perturbed-prompt output
excerpts.

\subsection{Clean Prompt}
\label{app:clean-prompt-example}

\begin{quote}
\small
You are preparing a quarterly supervisory review for Banco Norte, a retail
banking institution in Spain. Use the following current-quarter indicators:
CET1 ratio 10.8\%, liquidity coverage ratio 104\%, non-performing loan ratio
1.8\%, one-day 99\% VaR 4.5\% of trading capital, operational loss events
equal 0.4\% of gross income, and a short-term wholesale funding gap of 1.2\%
of liabilities. Interest-rate conditions are volatile. Assess systemic risk
across capital adequacy, liquidity, credit quality, market risk, operational
resilience, and interconnectedness. Provide a clear risk rating and specific
mitigation recommendations.
\end{quote}

\subsection{Perturbed Prompt Variants}
\label{app:perturbed-prompt-examples}

Table~\ref{tab:perturbed-prompt-examples} shows three perturbed versions of
the same clean prompt. Each variant combines a different stress vector with
semantic edits that inject contradiction, overload, ambiguity, and temporal
drift.

\begin{table}[h]
\centering
\small
\caption{Example perturbations for prompt \(\mathrm{P001}\).}
\label{tab:perturbed-prompt-examples}
\begin{tabular}{lrrrrp{7.2cm}}
\hline
Variant & Conflict & Load & Ambig. & Drift & Perturbation excerpt \\
\hline
\(\mathrm{P001\_V01}\) & 0.298 & 0.661 & 0.163 & 0.563 &
Adds an internal note stating that capital adequacy is both comfortably above
target and below the recovery trigger; adds unaudited ratios, net-of-guarantee
exposures, settlement-lag caveats, changed management overlays, excluded
hedging desks, and stale Q4 2025/Q2 2024 reporting windows. \\
\(\mathrm{P001\_V02}\) & 0.159 & 0.617 & 0.412 & 0.442 &
Replaces concrete references with ambiguous phrases such as ``that buffer''
and ``that exposure''; adds second-order effects from margin calls, drawdowns,
payment-system contagion, deposit beta, collateral haircuts, sovereign spread
widening, model-risk exceptions, and nine-month-old exposure data. \\
\(\mathrm{P001\_V03}\) & 0.152 & 0.494 & 0.500 & 0.580 &
Introduces ambiguous references such as ``that position'' and ``this area'';
adds a contradiction between contained volatility and severe mark-to-market
losses; adds scenario-comparison requirements, unaudited ratios, settlement
lags, and mismatched Q4 2025/Q2 2024 documents. \\
\hline
\end{tabular}
\end{table}

\subsection{Clean-Prompt Outputs}
\label{app:clean-output-examples}

Table~\ref{tab:clean-output-excerpts} shows representative excerpts from the
five architectures on the clean prompt. All architectures identify Banco Norte
as solvent but exposed to thin liquidity, volatile rates, and limited capital
headroom.

\begin{table}[h]
\centering
\small
\caption{Representative clean-prompt output excerpts.}
\label{tab:clean-output-excerpts}
\begin{tabular}{lp{10.4cm}}
\hline
Architecture & Output excerpt \\
\hline
A0 Flat &
``Overall risk rating: Moderate. Banco Norte appears broadly resilient but
shows elevated sensitivity to market and funding shocks given volatile rates,
a thin liquidity buffer, and reliance on wholesale funding.'' \\
A1 Hierarchical &
``Overall risk rating: Moderate-High. Banco Norte appears fundamentally
solvent, but its risk profile is fragile because several buffers are thin
rather than strong.'' \\
A2 Debate &
``Overall systemic risk rating: Moderate, with a negative outlook. Banco Norte
appears currently stable but not unambiguously comfortable.'' \\
A3 Meta-Adaptive &
``Overall systemic risk rating: Moderate / Amber. Banco Norte does not appear
immediately distressed, but several indicators are only modestly above or near
supervisory comfort zones.'' \\
A4 Ensemble &
``Weighted consensus risk rating: Moderate-High / Elevated Watch. The bank is
not in immediate distress, but its risk profile is fragile because liquidity is
thin and wholesale funding reliance is non-trivial.'' \\
\hline
\end{tabular}
\end{table}

\subsection{Perturbed-Prompt Outputs}
\label{app:perturbed-output-examples}

Table~\ref{tab:perturbed-output-excerpts} shows excerpts for
\(\mathrm{P001\_V01}\). The perturbed prompt introduces contradiction,
overload, ambiguity, and temporal drift. The outputs become more conservative
and devote more space to resolving inconsistent or stale evidence.

\begin{table}[h]
\centering
\small
\caption{Representative perturbed-prompt output excerpts for
\(\mathrm{P001\_V01}\).}
\label{tab:perturbed-output-excerpts}
\begin{tabular}{lp{10.4cm}}
\hline
Architecture & Output excerpt \\
\hline
A0 Flat &
``Overall risk rating: Moderate-to-High / Watchlist. Banco Norte appears
broadly solvent but increasingly vulnerable to a stress combination of capital
pressure, funding fragility, and asset-quality deterioration.'' \\
A1 Hierarchical &
``Overall risk rating: High / Elevated systemic risk. Banco Norte is not an
immediate capital failure case, but it shows a tight capital/liquidity position
with elevated tail risk across credit, market, liquidity, and operational
dimensions.'' \\
A2 Debate &
``Overall risk rating: Moderate-High. Banco Norte looks operationally viable
today, but the combination of borderline liquidity, moderate credit stress,
volatile interest rates, short-term wholesale funding dependence, and
data-quality caveats supports elevated supervisory concern.'' \\
A3 Meta-Adaptive &
``Supervisory stance: Heightened monitoring / negative bias. Banco Norte is
not in an immediate distress zone, but the combination of thin capital
headroom, marginal liquidity coverage, short-term wholesale reliance, and
data-quality/timeline inconsistencies makes the true risk profile uncertain.'' \\
A4 Ensemble &
``Overall risk rating: High / Elevated systemic vulnerability. Banco Norte
does not appear to be in immediate collapse, but it shows multiple thin
buffers, weak data reliability, and several reinforcing stress channels.'' \\
\hline
\end{tabular}
\end{table}

\subsection{Interpretation}
\label{app:example-interpretation}

The example illustrates the empirical pattern reported in the main text.
Table~\ref{tab:perturbed-prompt-examples} shows that the perturbed prompt does
not make the task easier: it introduces conflicting capital information,
additional caveats, ambiguous references, and stale reporting windows.
Tables~\ref{tab:clean-output-excerpts} and
\ref{tab:perturbed-output-excerpts} show that the architectures respond by
increasing caution and by surfacing different stress-handling strategies. This
is the qualitative counterpart of the CAFE measurement: stress reduces
immediate judged quality, but it also exposes architecture-specific variation
that can be reconstructed as observed effective stress.

\end{document}